\documentclass{article}

\usepackage[american]{babel}
\usepackage{csquotes}
\usepackage[style=authoryear-comp,citetracker=true,maxbibnames=99,maxcitenames=2,giveninits=true,backend=bibtex]{biblatex}
\DeclareNameAlias{sortname}{last-first}
\AtBeginBibliography{%
}
\renewbibmacro{in:}{}
\DeclareFieldFormat[article]{title}{#1}
\DeclareFieldFormat{pages}{#1}
\renewbibmacro*{volume+number+eid}{%
  \printfield{volume}%
  \setunit*{\addnbthinspace}
  \printfield{number}%
  \setunit{\addcomma\space}%
  \printfield{eid}}
\DeclareFieldFormat[article]{number}{\mkbibparens{#1}}
\DeclareFieldFormat[article]{volume}{\textit{#1}}

\DeclareFieldFormat{sentencecase}{\MakeSentenceCase{#1}}

\renewbibmacro*{title}{%
  \ifthenelse{\iffieldundef{title}\AND\iffieldundef{subtitle}}
    {}
    {\ifthenelse{\ifentrytype{article}\OR\ifentrytype{inbook}%
      \OR\ifentrytype{incollection}\OR\ifentrytype{inproceedings}%
      \OR\ifentrytype{inreference}\OR\ifentrytype{online}}
      {\printtext[title]{%
        \printfield[sentencecase]{title}%
        \setunit{\subtitlepunct}%
        \printfield[sentencecase]{subtitle}}}%
      {\printtext[title]{%
        \printfield[titlecase]{title}%
        \setunit{\subtitlepunct}%
        \printfield[titlecase]{subtitle}}}%
     \newunit}%
  \printfield{titleaddon}}

\AtEveryCitekey{\ifciteseen{}{\defcounter{maxnames}{99}}}
  
\usepackage{amsmath}
\usepackage{amsfonts}
\usepackage{amssymb}
\usepackage{amsthm}
\usepackage{chngcntr}
\usepackage{graphicx}
\usepackage{subcaption}
\usepackage{fancyhdr}
\usepackage{tabularx}
\usepackage{geometry}
\usepackage{setspace}
\usepackage[right]{eurosym}
\usepackage[printonlyused]{acronym}
\usepackage{floatflt}
\usepackage{colortbl}
\usepackage{paralist}
\usepackage{array}
\usepackage{titlesec}
\usepackage{parskip}
\usepackage[right]{eurosym}
\usepackage{longtable}
\usepackage[colorlinks=true,linkcolor=black,urlcolor=blue,citecolor=black,pdfpagelabels=true]{hyperref}
\usepackage{mathtools}
\usepackage{booktabs,caption}
\usepackage[flushleft]{threeparttable}
\usepackage{hyperref}
\usepackage{dsfont}
\usepackage{enumitem}
\usepackage[toc,page]{appendix}
\usepackage{rotating}
\usepackage{bbold}
\usepackage{scrextend}
\usepackage{verbatim}

\usepackage{listings}
\makeatletter
\def\l@lstlisting#1#2{\@dottedtocline{1}{0em}{1em}{\hspace{1,5em} Lst. #1}{#2}}
\makeatother

\newcommand\posscite[1]{\citeauthor{#1}'s (\citeyear{#1})}

\numberwithin{equation}{section}

\theoremstyle{plain}
\newtheorem{assumption}{Assumption}

\newtheorem{theorem}{Theorem}
\newtheorem{lemma}{Lemma}
\newtheorem{corollary}{Corollary}

\newcommand{\R}{R}
\begin{document}

\renewcommand*{\thefootnote}{\fnsymbol{footnote}}
\thispagestyle{empty}
\begin{center}
	\Large
	\textbf{Efficient Difference-in-Differences Estimation with High-Dimensional Common Trend Confounding}\\
\vspace*{2cm}
\begin{large}
	Michael Zimmert\footnote{michael.zimmert@unisg.ch, Michael Zimmert is employed and funded by the Swiss Institute of Empirical Economic Research (SEW) of the University of St. Gallen (HSG), Varnbüelstrasse 14, CH-9000 St.Gallen.}\\
	University of St. Gallen (HSG)
	\end{large}
\vspace*{1cm}
\begin{abstract}
This study considers various semiparametric difference-in-differences models under different assumptions on the relation between the treatment group identifier, time and covariates for cross-sectional and panel data. The variance lower bound is shown to be sensitive to the model assumptions imposed implying a robustness-efficiency trade-off. The obtained efficient influence functions lead to estimators that are rate double robust and have desirable asymptotic properties under weak first stage convergence conditions. This enables to use sophisticated machine-learning algorithms that can cope with settings where common trend confounding is high-dimensional. The usefulness of the proposed estimators is assessed in an empirical example. It is shown that the efficiency-robustness trade-offs and the choice of first stage predictors can lead to divergent empirical results in practice.    
\end{abstract}	
	
	\vspace*{5.5cm}
\end{center}
\vspace*{\fill}
\textbf{JEL classification:} C14, C21\\~\\
\textbf{Keywords:} Semiparametric difference-in-differences, machine-learning, semiparametric efficiency bound, high-dimensional data, employment protection.\\~\\
\textbf{Acknowledgements:} Financial support from the Swiss National Science Foundation (SNSF) is gratefully acknowledged. The study is part of the project "Causal Analysis with Big Data" of the Swiss National Research Program "Big Data" (NRP 75).\\
I want to thank Michael Lechner, Michael Knaus, Bryan Graham, Anthony Strittmatter, Beatrix Eugster and Petyo Bonev for competent advice and Franziska Zimmert for motivating the research question. I also thank Gabriel Okasa for some very useful pieces of R code and Daniel Goller, Carina Steckenleiter and Jana Mareckova for helpful remarks and comments. Further, I want to thank all participants of the European Causal Inference Meeting 2019 in Bremen for constructive feedback and fruitful discussions.
\pagebreak

\renewcommand*{\thefootnote}{\arabic{footnote}}
\setcounter{footnote}{0}
\setcounter{page}{1}
\doublespacing
\allowdisplaybreaks
\section{Introduction}\label{sec:intro}
In difference-in-differences identification is ensured by the fact that the subpopulation that will be exposed to the treatment (treatment group) and the subpopulation that will not be exposed to the treatment (control group) would have developed equally in the absence of treatment.\footnote{For textbook treatments see for example \textcite{Athey_Imbens_2017}, \textcite{Lechner_2010} or \textcite{Imbens_Wooldridge_2009}.} Identification often relies on the assumption that the common trend holds conditional on covariates. Any underlying factor differently shifting the potential outcomes under non-treatment for the treatment group and the control group needs to be controlled for. However, even if the researcher can credibly identify the factors that may lead to common trend confounding, it is still unclear in what form covariates should ultimately enter the statistical model for several reasons.\\
Crucially, the statistical model depends on assumptions about the relation between the treatment group identifier, time and observed covariates. With cross-sectional data covariates might be needed to account for imbalances between treatment and control group \textit{and} across time. With some notable exceptions (\cite{Lechner_2010}, \cite{Hong_2013}, \cite{Stuart_Huskamp_Duckworth_Simmons_Song_Chernew_Barry_2014}, \cite{Lu_Nie_Wager_2019}) most studies in semiparametric difference-in-differences exclude time-varying treatment group compositions and covariates (e.g., \cite{Heckman_Ichimura_Todd_1997}, \cite{Abadie_2005}, \cite{SantAnna_Zhao_2020}, \cite{Chang_2020}). This paper investigates semiparametric difference-in-differences models under various assumptions on how covariates, time and treatment group composition are related. Efficient influence functions are derived under more or less restrictive assumptions and for different sampling schemes. We present various identification and efficiency results for low-dimensional semiparametric difference-in-differences models. Our results are sensitive to assumptions about how covariates enter the model. In particular, our results hint at a trade-off between the strength of the assumptions the researcher is willing to impose on the model and the efficiency bound that can be achieved under such assumptions. Further, our results suggest that there are cases where we might want to include covariates even if not needed for identification, as they could increase the precision of some of the estimators. A comparison of the efficiency bounds for cross-section and panel data allows to draw interesting conclusions about the efficiency loss when panel data is not available. We therefore contribute to the literature on semiparametric efficiency in causal inference settings (e.g., \cite{Hahn_1998}, \cite{Firpo_2007}, \cite{Froelich_2007}, \cite{Chen_Hong_Tarozzi_2008}, \cite{Cattaneo_2010}, \cite{Graham_Pinto_Egel_2016}, \cite{Lee_2018}). Such an analysis is typically based on the approach developed by \textcite{Newey_1990,Newey_1994} and \textcite{Bickel_Klaassen_Ritov_Wellner_1993}. \textcite{Chamberlain_1987,Chamberlain_1992} contributes an alternative approach based on moment conditions. \textcite{Graham_2011} establishes an equivalence result between the moment condition based approach and the approach of \textcite{Bickel_Klaassen_Ritov_Wellner_1993} for the general missing data problem. In parallel work \textcite{SantAnna_Zhao_2020} also consider efficiency theory for semiparametric difference-in-differences problems to derive efficient score functions. Their results crucially rely on a relatively strong stationarity assumption and are included in this paper as a special case. We also notice that a previous version of the present paper was the first that proposed efficiency bounds for the semiparametric difference-in-differences problem using \posscite{Graham_2011} equivalence result. It turns out that for the panel case the moment conditions exhaust all the information in the identifying assumptions while for the cross-sectional case they do not.\footnote{See \textcite{Zimmert_2018} on this. We do not follow \posscite{Graham_2011} approach in this version. However, we consider the insufficiency of the first and second stage moment conditions to exhaust all information necessary to derive the efficiency bound for the cross-sectional difference-in-differences case as an interesting topic for further research.}\\
The efficient influence functions derived imply plug-in estimators that allow to combine semiparametric difference-in-differences models with very flexible first stage estimators. This is important because there might be many different covariates that are supposed to measure the same economic channel for common trend confounding and it might be unclear in what functional form the covariates should be included in the model. These issues might be especially prevalent in difference-in-differences models. Often covariates like geographic or industry classifications are available for different levels of aggregation -- making covariate selection an even more tedious task. For standard parametric models usually used for difference-in-differences estimation (e.g., \cite{Card_1990}, \cite{Card_Krueger_1994}, \cite{Eissa_Liebman_1996}) or semiparametric models with parametric or nonparametric first stages (e.g., \cite{Abadie_2005}, \cite{SantAnna_Zhao_2020}) a high-dimensional covariate space will cause the estimator to break down. Advances in the supervised machine-learning literature\footnote{For an overview see e.g. \textcite{Hastie_Tibshirani_Friedman_2009}.} showed an immense potential to approach this problem by choosing a data-driven trade-off between the covariate dimension and the sample size at hand and were successfully integrated in causal inference settings (e.g., \cite{Belloni_Chen_Chernozhukov_Hansen_2012}, \cite{Zhang_Zhang_2014}, \cite{vandeGeer_Buehlmann_Ritov_Dezeure_2014}, \cite{Belloni_Chernozhukov_Hansen_2014}, \textcite{Athey_Imbens_Wager_2018}). This paper builds on the generic 'double machine-learning' framework developped in \textcite{Chernozhukov_Chetverikov_Demirer_Duflo_Hansen_Newey_2017}. A major insight from this work is that 'single'-robust estimators based on the treatment mechanism (e.g., \cite{Horvitz_Thompson_1952}, \cite{Hirano_Imbens_Ridder_2003}, \cite{Hahn_Ridder_2013}) or outcome based models (e.g., \cite{Hahn_1998}) are inappropriate with machine-learning generated first stages while 'double'-robust estimators (\cite{Robins_Rotnitzky_Zhao_1994}, \cite{Scharfstein_Rotnitzky_Robins_1999}) maintain good statistical properties. \textcite{Chernozhukov_Chetverikov_Demirer_Duflo_Hansen_Newey_2017} show that the (rate) double robustness properties can be used such that there is no effect of first stage nuisance parameter estimation under relatively weak convergence conditions of the first stage parameters. We modify and extent the double machine-learning framework in this paper. In contrast to \textcite{Chernozhukov_Chetverikov_Demirer_Duflo_Hansen_Newey_2017}, this paper does not rely on high-level conditions based on Gateaux differentiation to verify the rate double robustness properties. Alternatively, we provide easy-to-check conditions that cover a broad range of scores typically used in the causal inference literature. Focusing on a specific class of widely used score functions allows to derive generalizable convergence condition requirements. Crucially, this substantially reduces the computational burden when deriving the asymptotic properties of an estimator and requires fewer regularity conditions such as the existence of the derivative or the interchangeability of the derivative and the expectation operator. Further, some of the derived efficient influence functions imply a new class of plug-in estimators whose convergence conditions depend on the existence of higher-order moments of the outcome. These results do not trivially follow from existing theory (e.g., \cite[Section 5]{Chernozhukov_Chetverikov_Demirer_Duflo_Hansen_Newey_2017}). Our theoretical results on double machine-learning is therefore also useful beyond the scope of difference-in-differences estimation. Additionally, they allow us to derive first stage convergence conditions for different semiparametric difference-in-differences estimators. This enables to incorporate sophisticated supervised machine-learning algorithms that can cope with settings where the dimension of the covariate space is high. Plug-in estimators that follow from the derived efficient influence functions are shown to achieve the low-dimensional variance lower bound. Our results also indicate that for some cases there is a trade-off between estimation robustness and efficiency. Throughout the evolvement of this paper other related but independent work on semiparametric difference-in-differences estimation with machine-learning appeared. \textcite{Chang_2020} also considers difference-in-differences estimation under the strong stationarity assumption by directly applying the double machine-learning results of \textcite{Chernozhukov_Chetverikov_Demirer_Duflo_Hansen_Newey_2017}. His estimator does generally not attain the semiparametric efficiency bound. \textcite{Lu_Nie_Wager_2019} propose estimators that are robust against the distortion of the stationarity assumption but focus on another parameter, consider alternative estimation methods and do not derive efficiency results.\\
Finally, we adopt the methods to a well-known application and investigate whether the efficiency-robustness trade-offs matter in practice and the value added when using machine-learning methods instead of standard parametric estimation methods.\\
The following section introduces the setting and required notation. The development of semiparametric theory for the difference-in-differences problem will be the starting point of our analysis in Section \ref{sec:ideff}. Section \ref{sec:estinf} presents our results on estimation and inference. To assess the usability of the proposed methods, they are applied to real world data in Section \ref{sec:app}. The last section concludes. Most technical proofs are relegated to the Appendix.
\section{Setting and notation}\label{sec:setnot}
Random variables like $A$ are denoted by capital letters. They have realizations $A=a$ that are in the support of the random variable $\mathcal{A}$. $A=a$ has density $f_A(a)$. If $A$ is discrete we write $Pr(A=a)=f_A(a)$ as a shorthand. The cumulative distribution function is given by $F_A(a)$. Let $B$ be another random variable. Then independence between two random variables $A$ and $B$ is denoted by $A\perp B$. The expectation operator is defined by $\mathbb{E}$ and $\text{Var}$ is used as a shorthand for the variance. For a generic function $g=g(B)=g_{A}(B)$ we use $A$ in the subscript to remind us of the mapping $g: b\mapsto a$. The $L_p$ norm is denoted by $\left\lVert g(B)\right\rVert_p$. As a special case we use $\sup_{b\in\mathcal{B}}\lvert g(b)\rvert$ and $\lVert g(B)\rVert_{\infty}$ interchangeably to denote the supremum of the function. The infinum is $\inf_{b\in\mathcal{B}}\lvert g(b)\rvert$. Let $\beta$ be some parameter then we denote $\dot{g}(B,\beta)=\frac{\partial g(B,\beta)}{\partial\beta}$. Generically, $C>0$ denotes a constant.\\
Let $D$, $T$ and $G_{\tau}$ be binary indicator variables such that $d,t,g_{\tau}\in\{0,1\}$ where $\tau\in\mathcal{T}$ and either $\mathcal{T}=\{(d,t)\}$ or $\mathcal{T}=\{d\}$. In particular, $D=1$ for observations that belong to the treatment group, $T=1$ for observations that are observed in period 1 and $G_{d,t}=1$ if $D=d$ and $T=t$ and 0 otherwise and $G_d=1$ if $D=d$ and 0 otherwise.\footnote{Obviously for the latter case $G_d=D$. We introduce this notation to formulate results as general as possible throughout our exposition.} Denote the outcome variable by $Y$ and some further observed variables by $X$. We follow the established literature (e.g., \cite{Roy_1951}, \cite{Rubin_1974}) and let $Y^d(t)$ be the potential outcome variable that contains the potentially unobserved realizations of $Y$ for the state $D=d$ and $T=t$. The exposition additionally relies on the definition of some conditional expectations. In particular, we have $m_Y(d,t,x)=\mathbb{E}\left[Y|D=d,T=t,X=x\right]$ with $m_Y(x)=\sum_{d=0}^1\sum_{t=0}^1(-1)^{d+t}m_Y(d,t,x)$. Similarly, for $\Delta Y=Y(1)-Y(0)$ we have $m_{\Delta Y}(d,x)=\mathbb{E}\left[Y(1)-Y(0)|D=d,X=x\right]$ with $m_{\Delta Y}(x)=m_{\Delta Y}(1,x)-m_{\Delta Y}(0,x)$. Additionally, suppose that $A$ is a binary variable. Then we generically define the probabilities $p_{A=a}(b)=Pr(A=a|B=b)$, $p_{A}(b)=Pr(A=1|B=b)$, $p_{A=a}=Pr(A=a)$ and $p_{A}=Pr(A=1)$. For example we have $p_{D=d,T=t}(x)=Pr(D=d,T=t|X=x)$ and $p_{DT}=Pr(D=1,T=1)$. Note that the definition of $G_{\tau}$ allows us to flexibly write for example $m_Y(d,t,x)=m_Y(G_{d,t}=1,x)$ and similarly for the other parameters.\\
In difference-in-differences settings the researcher is generally interested in identifying the parameter $\theta=\mathbb{E}\left[Y^1(1)-Y^0(1)|D=1,T=1\right]$. It can be described as an average treatment effect on the treated (ATET) because the parameter is defined for those who actually receive the treatment ($D=1$, $T=1$). An average population effect cannot be identified because this would require a subpopulation for which the treatment vanishes between period $T=0$ and $T=1$ (for a discussion on this see \cite{Lechner_2010}). We also note that under a strong stationarity conditions or when panel data is available \textcite{Abadie_2005} shows that $\mathbb{E}\left[Y^1(1)-Y^0(1)|D=1\right]$ is identified. Notice that, without further assumptions, this parameter is not an ATET but an average treatment effect for the treatment group. Intuitively, panel data or a stationarity assumption ensures that the composition of the treatment group does not depend on $T$ and so $\mathbb{E}\left[Y^1(1)-Y^0(1)|D=1,T=0\right]=\mathbb{E}\left[Y^1(1)-Y^0(1)|D=1,T=1\right]$. As the ATET is identified under all assumptions made in this paper, we focus on this parameter but indicate whenever the treatment group effect equals the ATET.
\section{Identification and efficiency bounds}\label{sec:ideff}
\subsection{Repeated cross-sections}
\begin{assumption}[Data-generating process CS]\label{ass:dgpcs} Let $W(t)=(Y(t),D(t),T=t,X(t))$. (i) The i.i.d. sample of two repeated cross-sections with $W=\{W(0),W(1)\}=(Y,D,T,X)$ with observations $i=1,...,N$ is observed; (ii) The joint distribution $F_{W(0),W(1)}(w(0),w(1))=F_W(w)$ exists.
\end{assumption}
Assumption \ref{ass:dgpcs} describes the data-generating process (DGP) for the repeated cross-sections. It guarantees that we can use the pseudo-sample $W_i$ with observations $i=1,...,N$ and emphasizes that we have to cope with a merged sample problem where the sample sizes of $W(0)$ and $W(1)$, $N(0)$ and $N(1)$ obey $\frac{N(0)}{N(1)}\rightarrow C$ (\cite{Abadie_Imbens_2006}, \cite{Graham_Pinto_Egel_2016}). Also notice that the existence of the joint distribution implies that the dimension of $X$ is fixed. We will relax this condition in Section \ref{sec:estinf}.\\
In what follows we discriminate five settings (CS-1) to (CS-5) that describe different assumption sets on the relation between $D$, $T$ and $X$. In (CS-1) we do not make any further assumptions. It is the fully robust setting. Settings (CS-2)-(CS-5) are comprised in Assumption \ref{ass:reldtxcs}.
\begin{assumption}[Relation of $D$, $T$ and $X$]\label{ass:reldtxcs}
The variables $D$, $T$ and $X$ are assumed to be related in the following ways. (CS-2) Conditional independence of $D$ and $T$, $D\perp T|X=x$ for every $x\in\mathcal{X}$; (CS-3) Independence of $X$ and $T$, $X\perp T$; (CS-4) Joint independence of $D$ and $X$ from $T$, $(D,X)\perp T$; (CS-5) Mutual independence of $D$, $T$ and $X$, $(D\perp T\perp X)$.
\end{assumption}
(CS-2) allows for time varying $D$ and $X$ but requires that all time variation in $D$ is fully captured by $X$. (CS-3) does not allow for time-varying $X$ but $D$ may still follow a time trend. (CS-4) implies the strong stationarity assumption used by \textcite{Abadie_2005}, \textcite{SantAnna_Zhao_2020} and \textcite{Chang_2020}. It excludes time variation in $D$ and $X$. (CS-5) is the `experimental' setting. Even though the outcome might depend on $X$, there are no imbalances neither between treatment and control group nor across time.\\
Since Assumption \ref{ass:reldtxcs} only contains conditions on the relation of observed random variables, it is in principal testable. Especially, a significant correlation between $D$ and $T$ rules out settings (CS-4) and (CS-5). Also notice that (CS-2) and (CS-3) are mutually exclusive and that the restrictiveness of the assumptions can be ordered as (CS-5), (CS-4), (CS-3)/(CS-2), (CS-1).
\subsubsection{Identification}
\begin{assumption}[Identification CS]\label{ass:idcs}
For any $d,t\in\{0,1\}$ and $x\in\mathcal{X}$,
\begin{itemize}
\item[(i)] (Observational Rule) For each observation $i$, the outcome $Y_i=\sum_{d}\sum_{t}G_{{d,t}_i}Y_i^d(t)$ is observed;
\item[(ii)] (Common Support) The propensity score $p_{D=d,T=t}(x)$ is bounded away from zero;
\item[(iii)] (No Anticipation) $\mathbb{E}\left[Y^1(0)-Y^0(0)|D=1,T=0,X=x\right]=0$;
\item[(iv)] (Conditional Common Trends)
\begin{align*}
&\mathbb{E}\left[Y^0(1)|D=0,T=1,X=x\right]-\mathbb{E}\left[Y^0(0)|D=0,T=0,X=x\right]\\
&=\mathbb{E}\left[Y^0(1)|D=1,T=1,X=x\right]-\mathbb{E}\left[Y^0(0)|D=1,T=0,X=x\right].
\end{align*}
\end{itemize}
\end{assumption}
Assumption \ref{ass:idcs} yields an identification result for $\theta$ with cross-sectional data. The Observational Rule underscores that we only observe $Y$ and not $Y(0)$ and $Y(1)$ for every observation. It rules out that observations in the treatment group can be part of the control group or that observations in period $T=0$ are again observed in $T=1$. Notice that since we consider a pseudo-sample, this does not rule out individuals from an actual population from being re-sampled in the second cross-section. Common Support is necessary to guarantee the existence of conditional expectations. Since we are only interested in the ATET, the propensity score only needs to be bounded away from zero for identification. The No Anticipation condition rules out an effect of the treatment in period $T=0$ for the treatment group. The Conditional Common Trends condition requires that conditional on the covariates the treatment and the control group would have developed equally in the absence of the treatment. To allow for a more compact representation of the results, some further notation is introduced. Let $q_{CS;D=d,T=t}(X)$ denote the conditional probability function $p_{D=d,T=t}(X)$ under some of the specific assumptions on the relation of $D$, $T$ and $X$ in (CS-1) to (CS-5). For example we have $q_{CS-2;D=1,T=1}(X)=p_D(X)p_T(X)$. Equivalently, denote by $q_{CS;DT}$ the unconditional probability $p_{DT}$ under some specific assumption (CS-1) to (CS-5).
\begin{lemma}\label{lm:idcs}
Under Assumptions \ref{ass:dgpcs} and \ref{ass:idcs} the parameter $\theta=\mathbb{E}\left[Y^1(1)-Y^0(1)|D=1,T=1\right]$  is identified as $\mathbb{E}\left[m_Y(X)\frac{q_{CS;D=1,T=1}(X)}{q_{CS;DT}}\right]$.
\end{lemma}
\textit{Proof:} Notice that
\begin{align*}
\mathbb{E}\left[Y^1(1)-Y^0(1)|D=1,T=1,X=x\right]&=m_Y(1,1,x)-\mathbb{E}\left[Y^0(0)|D=1,T=0,X=x\right]\\
&-\mathbb{E}\left[Y^0(1)|D=0,T=1,X=x\right]+\mathbb{E}\left[Y^0(0)|D=0,T=0,X=x\right]\\
&=m_Y(1,1,x)-\mathbb{E}\left[Y^1(0)|D=1,T=0,X=x\right]\\
&-m_Y(0,1,x)+m_Y(0,0,x)\\
&=m_Y(X)
\end{align*}
using Assumptions \ref{ass:idcs}. Further, for the conditional density function $f_{X|D=d,T=t}(x|d,t)$
\begin{align*}
\theta=\int m_Y(X)f_{X|D=1,T=1}(x|1,1)dx=\int m_Y(X)\frac{p_{D=1,T=1}(x)}{p_{DT}}f_X(x)dx=\int m_Y(X)\frac{q_{CS;D=1,T=1}(x)}{q_{CS;DT}}f_X(x)dx.
\end{align*}
For cases (CS-4) and (CS-5) the treatment group effect is identified.
\subsubsection{Semiparametric efficiency bounds}
\begin{theorem}[Semiparametric efficiency bounds CS]\label{thm:effcs}
Suppose that Assumptions \ref{ass:dgpcs} and \ref{ass:idcs} hold. Then under each of the settings in (CS-1)-(CS-5) the efficient influence function is given by
\begin{align*}
\psi^{*}_{CS}(W;\theta)=\frac{q_{CS;D=1,T=1}(X)}{q_{CS;DT}}\psi^{*a}_{CS}(W)+\psi^{*b}_{CS}(W)\left(m_Y(X)-\theta\right)
\end{align*}
where
\begin{align*}
\psi^{*a}_{CS}(W)&=\sum_{d=0}^1\sum_{t=0}^1(-1)^{(d+t)}\frac{G_{d,t}}{q_{CS;D=d,T=t}(X)}\left(Y-m_Y(d,t,X)\right)
\end{align*}
and $\psi^{*b}_{CS-1}(W)=\frac{DT}{p_{DT}}$, $\psi^{*b}_{CS-2}(W)=\frac{Dp_T(X)+p_D(X)T-p_D(X)p_T(X)}{p_{DT}}$, $\psi^{*b}_{CS-3}(W)=\frac{T(D-p_D(1,X))+p_D(1,X)p_T}{p_{D}(1)p_T}$, $\psi^{*b}_{CS-4}(W)=\frac{D}{p_D}$ and $\psi^{*b}_{CS-5}(W)=1$. The semiparametric efficiency bound for settings (CS-1) to (CS-5) is $\mathbb{E}\left[\psi^{*}_{CS}(W;\theta)^2\right]$. 
\end{theorem}
\textit{Proof:} see Appendix \ref{app:effcs}.\\
Theorem \ref{thm:effcs} is our first main result. Under each of the settings (CS-1)-(CS-5) an influence function with a different adjustment term $\psi^{*b}_{CS}(W)$ is derived implying different efficiency bounds for the different settings. An important implication is summarized in Corollary \ref{cor:releffcs1}. 
\begin{corollary}\label{cor:releffcs1}
In terms of the asymptotic variance bound, the value of knowing that one of the assumed relations of $D$, $T$ and $X$ in (CS-2) to (CS-5) holds relative to (CS-1) is given by
\begin{align*}
\Delta_{CS-1,CS-2}&=\mathbb{E}\left[\frac{p_D(X)^2p_T(X)^2}{p_{DT}^2}\left(m_Y(X)-\theta\right)^2\left(\frac{1}{p_D(X)p_T(X)}-\frac{1}{p_D(X)}-\frac{1}{p_T(X)}+1\right)\right]\\
\Delta_{CS-1,CS-3}&=\mathbb{E}\left[\frac{p_D(1,X)^2}{p_{D}(1)^2}\left(m_Y(X)-\theta\right)^2\left(\frac{1}{p_T}-1\right)\right]\\
\Delta_{CS-1,CS-4}&=\mathbb{E}\left[\frac{p_D(X)^2}{p_{D}^2}\left(m_Y(X)-\theta\right)^2\frac{1}{p_D(X)}\left(\frac{1}{p_T}-1\right)\right]\\
\Delta_{CS-1,CS-5}&=\mathbb{E}\left[\left(m_Y(X)-\theta\right)^2\left(\frac{1}{p_Dp_T}-1\right)\right].
\end{align*}
\end{corollary}
\textit{Proof:} see Appendix \ref{app:releffcs1}.\\
Unambiguously, $\Delta_{CS-1,CS-2},\Delta_{CS-1,CS-3},\Delta_{CS-1,CS-4},\Delta_{CS-1,CS-5}>0$. Hence, knowing that one of the assumptions (CS-2)-(CS-5) is true results in lower efficiency bounds compared to not making any assumptions (CS-1) about the relation between $D$, $T$ and $X$. Since (CS-2) and (CS-3) contain (CS-4) and (CS-4) contains (CS-5) similar results can be shown for these cases. To conclude, more restrictive assumptions about the relation between $D$, $T$ and $X$ imply lower efficiency bounds. This hints at a trade-off regarding the robustness with respect to the model assumptions imposed and the variance lower bound that is asymptotically achievable.
\subsection{Panel data}
\begin{assumption}[Data-generating process PA]\label{ass:dgppa}
(i) The i.i.d. sample of a two-period panel with $W=(Y(0),Y(1),D,X)$ and observations $i=1,...,N$ is observed; (ii) The distribution $F_{W}(w)$ exists.
\end{assumption}
Assumption \ref{ass:dgppa} describes the DGP when panel data is available. In contrast to Assumption \ref{ass:dgpcs}, the sample is not merged and contains $i=1,...,N$ unique individuals. Also the outcomes $Y(0)$ and $Y(1)$ are directly observable in the sample for one observation and thus do not need to be inferred.\\
For panel data we discriminate two settings (PA-1) and (PA-2) that describe different assumption sets on the relation between $D$ and $X$. Since we do not need a time indicator to describe the sample, the analysis is limited to the relation between $D$ and $X$. In (PA-1) we do not make any further assumptions. It is the fully robust setting. 
\begin{assumption}[Relation of $D$ and $X$]\label{ass:reldxpa}
The variables $D$ and $X$ are assumed to be independent, $D\perp X$ (PA-2).
\end{assumption}
(PA-2) is more restrictive in the sense that imbalances between the treatment and the control group are ruled out.
\subsubsection{Identification}
\begin{assumption}[Identification PA]\label{ass:idpa}
For any $d,t\in\{0,1\}$ and $x\in\mathcal{X}$, 
\begin{itemize}
\item[(i)] (Observational Rule) For each observation $i$, the outcomes\\ $Y_i(t)=D_iY_i^1(t)+(1-D_i)Y_i^0(t)$ are observed;
\item[(ii)] (Common Support) The propensity score $p_{D}(x)$ is bounded away from zero;
\item[(iii)] (No Anticipation) $\mathbb{E}\left[Y^1(0)-Y^0(0)|D=1,X=x\right]=0$;
\item[(iv)] (Conditional Common Trends) 
\begin{align*}
\mathbb{E}\left[Y^0(1)-Y^0(0)|D=0,X=x\right]=\mathbb{E}\left[Y^0(1)-Y^0(0)|D=1,X=x\right].
\end{align*}
\end{itemize}
\end{assumption}
Similarly to Assumption \ref{ass:idcs}, Assumption \ref{ass:idpa} ensures the identification of $\theta$ when panel data is available. The Observational Rule guarantees that for both outcomes $Y(0)$ and $Y(1)$ an observation cannot be part of the treatment and the control group at the same time. The other assumptions are adapted in a straightforward manner from the cross-sectional setting.\\
Again we let $q_{PA;D=d}(X)$ denote the conditional probability function $p_{D=d}(X)$ under either (PA-1) or (PA-2).
\begin{lemma}\label{lm:idpa}
Under Assumptions \ref{ass:dgppa} and \ref{ass:idpa} the parameter $\theta=\mathbb{E}\left[Y^1(1)-Y^0(1)|D=1\right]$  is identified as $\mathbb{E}\left[m_{\Delta Y}(X)\frac{q_{PA;D}(X)}{p_{D}}\right]$.
\end{lemma}
\textit{Proof:} The proof follows similar to Lemma \ref{lm:idcs}.\\
In the panel the treatment group effect is identified.
\subsubsection{Efficiency bounds}
\begin{theorem}[Semiparametric efficiency bounds PA]\label{thm:effpa}
Suppose that Assumptions \ref{ass:dgppa} and \ref{ass:idpa} hold. Then under each of the settings (PA-1) and (PA-2) the efficient influence function is given by
\begin{align*}
\psi^{*}_{PA}(W;\theta)=\frac{q_{PA;D=1}(X)}{p_D}\psi^{*a}_{PA}(W)+\psi^{*b}_{PA}(W)\left(m_{\Delta Y}(X)-\theta\right)
\end{align*}
where
\begin{align*}
\psi^{*a}_{PA}(W)&=\frac{D}{q_{PA;D=1}(X)}\left(Y(1)-Y(0)-m_{\Delta Y}(1,X)\right)-\frac{1-D}{q_{PA;D=0}(X)}\left(Y(1)-Y(0)-m_{\Delta Y}(0,X)\right)
\end{align*}
and $\psi^{*b}_{PA-1}(W)=\frac{D}{p_{D}}$ and $\psi^{*b}_{PA-2}(W)=1$. The semiparametric efficiency bound for settings (PA-1) and (PA-2) is $\mathbb{E}\left[\psi^{*}_{PA}(W;\theta)^2\right]$. 
\end{theorem}
\textit{Proof:} see Appendix \ref{app:effpa}.\\
Theorem \ref{thm:effpa} is our second main result. Under (PA-1) and (PA-2) an influence function with a different adjustment term $\psi^{*b}_{PA}(W)$ is derived implying different efficiency bounds for (PA-1) and (PA-2). An important implication is summarized in Corollary \ref{cor:releffpa1}.
\begin{corollary}\label{cor:releffpa1}
In terms of the asymptotic variance bound, the value of knowing that $D\perp X$ is given by
\begin{align*}
\Delta_{PA-1,PA-2}=\mathbb{E}\left[\left(m_{\Delta Y}(X)-\theta\right)^2\left(\frac{1}{p_D}-1\right)\right]
\end{align*}
\end{corollary}
\textit{Proof:} The proof follows similarly to Corollary \ref{cor:releffcs1}.\\
Again, since $\Delta_{PA-1,PA-2}>0$, the more restrictive assumption (PA-2) is associated with a lower efficiency bound. Hence, the robustness-efficiency trade-off also materializes for panel data.\\
Having derived results for both cross-sectional and panel data, it might be of interest to compare the variance lower bounds under both sampling schemes. Corollary \ref{cor:releffcspa} summarizes the implications of Theorems \ref{thm:effcs} and \ref{thm:effpa} for the relative efficiency between panel and cross-sectional data.
\begin{corollary}\label{cor:releffcspa}
In terms of the asymptotic variance bound, the value of knowing the panel structure under no further assumptions is given by
\begin{align*}
\Delta_{CS-1,PA-1}=\mathbb{E}\left[\frac{p_D(X)^2}{p_D^2}\left(\sum_{d=0}^1\frac{\text{Var}(Y(1)+Y(0)|D=d,X)}{p_{D=d}(X)}+\frac{(m_{\Delta Y}(X)-\theta)^2}{p_D(X)}\right)\right].
\end{align*}
The minimum value of knowing the panel structure is given by
\begin{align*}
\Delta_{CS-5,PA-1}=\mathbb{E}\left[\sum_{d=0}^1\frac{\text{Var}(Y(1)+Y(0)|D=d,X)}{p_{D=d}}+(m_{\Delta Y}(X)-\theta)^2\left(1-\frac{1}{p_D}\right)\right].
\end{align*}
\end{corollary}
\textit{Proof:} see Appendix \ref{app:releffcspa}.\\
Since $\Delta_{CS-1,PA-1}>0$, under no further condition on the relation between $D$, $T$ and $X$ the first result of the corollary shows that knowing the panel structure generally reduces the variance lower bound that is asymptotically achievable under comparable assumptions. This is an intuitive result because the information in a panel is unambiguously richer. $\Delta_{CS-1,PA-1}$ can therefore also be seen as the gain from the potentially more costly panel sampling scheme. Since the difference $\Delta Y$ is observed with panel data, the variance lower bound only contains propensity score reweighted conditional variances. For cross-sectional data the conditional variances are additionally reweighted by $p_{T=t}$ resulting in an efficiency loss. Further, notice that the gain from observing the panel $\Delta_{CS-1,PA-1}$ is higher (lower) when the correlation between $Y(0)$ and $Y(1)$ is positive (negative). This can be explained by the fact that the variance of $\Delta Y$ is lowest when $Y(0)$ and $Y(1)$ are positively correlated. Hence, the panel becomes more valuable when the observed difference $\Delta Y$ is less volatile. The second result of the corollary assesses the value of making assumptions relative to having access to panel versus cross-sectional data. It is hypothesized that the researcher knows that (CS-5) is correct and has panel data available where he uses the stronger than necessary setting (PA-1). From Theorem \ref{thm:effcs} and Corollary \ref{cor:releffcs1} we know that among all settings (CS-1)-(CS-5) the minimum variance lower bound with cross-sectional data is achieved when making assumption (CS-5). From Theorem \ref{thm:effpa} and Corollary \ref{cor:releffpa1} we know that among (PA-1) and (PA-2) the maximum variance lower bound with panel data is achieved when making assumption (PA-1). The difference among these settings when panel data becomes available $\Delta_{CS-5,PA-1}$ thus measures  the value of making assumptions relative to having access to better data. Notice that $\Delta_{CS-5,PA-1}\lesseqgtr 0$. Hence, having access to panel data does not necessarily lead to lower efficiency bounds. Rather, the corollary shows the importance of making adequate assumptions on the relation between $D$, ($T$) and $X$.
\section{Estimation and Inference}\label{sec:estinf}
\subsection{High-dimensional data and important building blocks}
Denote by $\eta$ a set of nuisance parameters that are generally unknown and have to be estimated in a first stage. From Section \ref{sec:ideff}, $\eta$ consists of functions of $X$. By Assumptions \ref{ass:dgpcs} and \ref{ass:dgppa} the results of Section \ref{sec:ideff} only hold when the dimension of the covariates $\lambda_X$ is fixed. For estimation and inference we relax this condition and assume that $X\in\mathbb{R}^{\lambda_X}$ and potentially $\lambda_X\rightarrow\infty$ when $N\rightarrow\infty$. Conditions in the next subsection will describe the concrete growth rates of $\lambda_X$ in relation to $N$.\\
Let $\psi$ be a function of the observed variables $W$ and the nuisance parameters. $W$ contains the generic outcome variable $\tilde{Y}$ and $\eta$ contains some projection on $\tilde{Y}$ denoted by $m_{\tilde{Y}}(\cdot)$. In contrast to Section \ref{sec:ideff}, we now explicitly indicate that the function $\psi$ depends on $\eta$ and consider $\psi$ to be of the general form
\begin{align*}
\psi(W,\eta;\theta)=\psi(W,\eta)-\psi^b(W,\eta)\theta=\frac{q_{1}(X)}{q_1}\psi^a(W,\eta)+\psi^b(W,\eta)(m_{\tilde{Y}}(X)-\theta)
\end{align*}
In particular, the function $\psi(W,\eta)$ is a sum over the index $\tau$ of terms in the form 
\begin{align*}
\frac{q_1(X)}{q_1}\frac{G_{\tau}}{q_{G_\tau}(X)}\left(\tilde{Y}-m_{\tilde{Y}}(G_{\tau}=1,X)\right)+\psi^b(W,\eta)m_{\tilde{Y}}\left(G_{\tau}=1,X\right)
\end{align*}
where $q_{G_{\tau}}(x)=Pr(G_{\tau}=1|X=x)$ under some specific assumption on the relation between $G_{\tau}$ and $X$. $q_1(X)$ and $q_1$ are shorthand symbols for $q_{G_{\tau}}(X)$ and $q_{G_{\tau}}$ with $\tau=(1,1)$ in the cross-sectional and $\tau=1$ in the panel case. We assume that $\mathbb{E}\left[\psi(W,\eta;\theta)\right]=0$ such that $\theta$ is identified as
\begin{align*}
\theta=\frac{\mathbb{E}\left[\frac{q_1(X)}{q_1}\psi^a(W,\eta)+\psi^b(W,\eta)m_{\tilde{Y}}(X)\right]}{\mathbb{E}\left[\psi^b(W,\eta)\right]}.
\end{align*}
A plug-in estimator uses the ratio of sample averages with estimated nuisance parameters. Following the suggestions of \textcite{Chernozhukov_Chetverikov_Demirer_Duflo_Hansen_Newey_2017} we apply a cross-fitting algorithm for the nuisance parameter estimation step in order to guarantee that the resulting estimators of $\psi(W,\eta)$ and $\psi^b(W,\eta)$ consist of independent observations. The details of the estimation strategy are outlined in the algorithm below.
\begin{figure}[h]
\centering
\onehalfspacing
\fbox{\begin{minipage}{0.95\textwidth}
\onehalfspacing
\textbf{Cross-fitting algorithm:}\\~\\
Suppose that the set of random variables $W$ can be indexed by $i$ such that the sample is described by $W_{i}$ for $i=1,...,N$. Randomly split the sample in $K$ equal subsamples of size $n=\frac{N}{K}$. For each of the subsamples with index $k=1,...,K$ define the set of sample indices in subsample $k$ by $\mathcal{I}^k$ and the set of sample indices not in $k$ by $\mathcal{I}^{-k}$. Then a cross-fitted estimator $\hat{\theta}$ is obtained by the following procedure.\\
\textbf{for} $k=1$ \textbf{to} $K$\textbf{:}
\begin{enumerate}
\item Estimate all nuisance parameters $\eta$ using $W_{i\in\mathcal{I}^{-k}}$ and define these estimators as $\hat{\eta}_{-k}$.
\item Use $W_{i\in\mathcal{I}^k}$ to obtain $\frac{1}{n}\sum_{i\in\mathcal{I}^k}^n\psi(W_i,\hat{\eta}_{-k})$ and $\frac{1}{n}\sum_{i\in\mathcal{I}^k}^n\psi^b(W_i,\hat{\eta}_{-k})$.
\end{enumerate}
\textbf{endfor.}\\
Finally, construct the estimator $\hat{\theta}=\frac{\frac{1}{N}\sum_{k=1}^K\sum_{i\in\mathcal{I}^k}^n\psi(W_i,\hat{\eta}_{-k})}{\frac{1}{N}\sum_{k=1}^K\sum_{i\in\mathcal{I}^k}^n\psi^b(W_i,\hat{\eta}_{-k})}$.
\end{minipage}}
\end{figure}
Our conditions on the first stage nuisance parameter convergence rates should cover a wide range of estimators. We therefore rely on $L_2$ convergence rates. To reduce the notational burden, generically write $L_2$-rates for cross-fitted nuisance parameters as $\epsilon_{\eta}=\left\lVert\hat{\eta}_{-k}-\eta\right\rVert_2$.
\subsection{Asymptotic results}
In order to derive asymptotic results for the cross-fitting estimator described in the previous subsection, we have to make several assumptions.
\begin{assumption}[Existence of higher-order moments]\label{ass:estinfy}
(i) For $r\in\mathbb{N}$ the first $r\geq 2$ moments of $\tilde{Y}$ exist; (ii) For all $x\in\mathcal{X}$, $\tau\in\mathcal{T}$, the conditional variance $\text{Var}\left(\tilde{Y}|G_{\tau}=1,X=x\right)$ exists.
\end{assumption}
Notice that Assumption \ref{ass:estinfy} at least requires the existence of the second moment of the outcome. We will later show that for some difference-in-differences estimators the existence of higher-order moments interacts with the requirements of first stage convergence conditions.
\begin{assumption}[Boundedness of propensity scores]\label{ass:estinfp}
For all $\tau\in\mathcal{T}$, conditional probabilities and their estimators obey $0<\inf_{x\in\mathcal{X}}\lvert q_{G_{\tau}}(x)\rvert<\sup_{x\in\mathcal{X}}\lvert q_{G_{\tau}}(x)\rvert<1$ and $0<\inf_{x\in\mathcal{X}}\lvert \hat{q}_{G_{\tau}}(x)_{-k}\rvert<\sup_{x\in\mathcal{X}}\lvert \hat{q}_{G_{\tau}}(x)_{-k}\rvert<1$.
\end{assumption}
Assumption \ref{ass:estinfp} is stronger than the usual Common Support conditions made in Assumptions \ref{ass:idcs} and \ref{ass:idpa} to obtain the identification results in Section \ref{sec:ideff}. All conditional probabilities and their estimators are strictly bounded away from zero and one. The assumption precludes using for example the Linear Probability Model or the linear Lasso for estimating the propensity scores.
\begin{assumption}[Behaviour of adjustment term]\label{ass:estinfb}
The term $\psi^b(W,\eta)$ obeys,
\begin{itemize}
\item[(i)] $\mathbb{E}\left[\psi^b(W,\eta)|X\right]=\frac{q_1(X)}{q_1}$;
\item[(ii)] $0<\inf\lvert\frac{1}{N}\sum_{k=1}^K\sum_{i\in\mathcal{I}^k}^n\psi^b(W_i,\eta)\rvert\leq C$, $0<\inf\lvert \frac{1}{N}\sum_{k=1}^K\sum_{i\in\mathcal{I}^k}^n\psi^b(W_i,\hat{\eta}_{-k})\rvert\leq C$;
\item[(iii)] for all $\tau\in\mathcal{T}$, $\left\lVert\frac{G_{\tau}}{q_1}\frac{q_1(X)}{q_{G_{\tau}}(X)}-\psi^b(W,\eta)\right\rVert_{\infty}\leq C$.
\end{itemize}
\end{assumption}
Assumption \ref{ass:estinfb} contains some further regularity conditions on the behaviour of the term $\psi^b(W,\eta)$ that are easily satisfied for the difference-in-differences estimators to be considered. Condition (i) is in principle redundant because if it would not hold then $\mathbb{E}\left[\psi(W,\eta;\theta)\right]=0$ would not be satisfied. We keep it, however, to remind us that the term cannot be of arbitrary form. Since the cross-fitted estimator $\hat{\theta}$ involves dividing by the sample plug-in average of $\psi^b(W,\eta)$, conditions (ii) are needed to guarantee that $\hat{\theta}$ is nicely behaved. The last condition is satisfied whenever the infinum of $\psi^b(W,\eta)$ is zero.
\begin{assumption}[First stage convergence conditions]\label{ass:estinfjoint}
The convergence conditions
\begin{itemize} 
\item[(i)] $\left\lVert \psi^b(W,\hat{\eta}_{-k})-\psi^b(W,\eta)\right\rVert_{2}=o_p(1)$ and $\left\lVert \psi^b(W,\hat{\eta}_{-k})-\psi^b(W,\eta)\right\rVert_{\infty}=O_p(1)$;
\item[(ii)] $\mathbb{E}\left[\psi^b(W,\hat{\eta}_{-k})-\psi^b(W,\eta)\right]=o_p\left(N^{-\frac{1}{2}}\right)$;
\item[(iii)] $\mathbb{E}\left[\left(\psi^b(W,\hat{\eta}_{-k})-\psi^b(W,\eta)\right)\hat{m}_{\tilde{Y}}(G_{\tau}=1,X)\right]=o_p\left(N^{-\frac{1}{2}}\right)$ and\\
$\left\lVert\left(\psi^b(W,\hat{\eta}_{-k})-\psi^b(W,\eta)\right)m_{\tilde{Y}}(G_{\tau}=1,X)\right\rVert_2=o_p(1)$;
\item[(iv)] $\left\lVert\hat{m}_{\tilde{Y}}(G_{\tau}=1,X)_{-k}-m_{\tilde{Y}}(G_{\tau}=1,X)\right\rVert_2=o_p(1)$ and $\left\lVert\frac{\hat{q}_1(X)_{-k}}{\hat{q}_{G_{\tau}}(X)_{-k}}-\frac{q_1(X)}{q_{G_{\tau}}(X)}\right\rVert_2=o_p(1)$;
\item[(v)] $\left\lVert\frac{\hat{q}_1(X)_{-k}}{\hat{q}_{G_{\tau}}(X)_{-k}}-\frac{q_1(X)}{q_{G_{\tau}}(X)}\right\rVert_2\times\left\lVert\hat{m}_{\tilde{Y}}(G_{\tau}=1,X)_{-k}-m_{\tilde{Y}}(G_{\tau}=1,X)\right\rVert_2=o_p\left(N^{-\frac{1}{2}}\right)$
\end{itemize}
are satisfied for all $\tau\in\mathcal{T}$.
\end{assumption}
Assumption \ref{ass:estinfjoint} comprises the required coupled convergence conditions that are at the centre of our theoretical argument. The assumption ensures that the first stage nuisance parameter estimation has no effect on the asymptotic behaviour of the cross-fitted estimator. Conditions (i)-(iii) contain the convergence requirements requirements for term $\psi^b(W,\eta)$. Notice that they are trivially satisfied whenever $\psi^b=\psi^b(W)$. Hence, when the term does not depend on nuisance parameters but just on observed variables, conditions (i)-(iii) always hold. However, some of the efficient influence functions considered for the cross-sectional case in Section \ref{sec:ideff} suggest that one may also use functions where $\psi^b=\psi^b(W,\eta)$. This represents the first extension to the work of \textcite{Chernozhukov_Chetverikov_Demirer_Duflo_Hansen_Newey_2017} who just consider problems of the latter type when deriving first stage convergence rate conditions. Condition (iv) requires that for all $\tau\in\mathcal{T}$ the nuisance parameters converge in $L_2$. Condition (v) requires that for all $\tau\in\mathcal{T}$ the conditional probability and the outcome nuisances jointly achieve $\sqrt{N}$-convergence. Conditions (iv) and (v) are easy-to-check conditions because they can be applied to all sorts of reweighting schemes. In particular, the specific convergence conditions for all $\tau\in\mathcal{T}$ can be directly retrieved from the conditions provided. This represents the second extension to the work of \textcite{Chernozhukov_Chetverikov_Demirer_Duflo_Hansen_Newey_2017} who rely on Gateaux differentiation and implied regularity conditions to derive required convergence rate for specific settings. Our theory covers all functions $\psi(W,\eta;\theta)$ where $\psi(W,\eta)$ can be written as a sum over some index $\tau$. This includes the usual parameters of interest and settings in causal econometrics such as selection-on-observables with binary or multiple treatments, difference-in-differences and instrumental variables.\\
The $L_2$ rate conditions in Assumption \ref{ass:estinfjoint} (iv) and (v) can be shown to be satisfied for many supervised machine-learning algorithms under sparsity conditions. For example \textcite{Belloni_Chernozhukov_2013} show that the predictive error of the Lasso is of order $O_p\left(\sqrt{\frac{s\log\max(\lambda_X,N)}{N}}\right)$ where $s$ is the unknown number of true coefficients in the oracle model. Assumption $\ref{ass:estinfjoint}$ (iv) and (v) then require that $\frac{s^2\log^2\max(\lambda_X,N)}{N}\rightarrow 0$ (if $s$ is the same in the propensity score and the outcome nuisance models). It follows that $\lambda_X\rightarrow\infty$ when $N\rightarrow\infty$ under the sparsity condition. Thus, the dimension of the covariates can be high in the sense that it can (slowly) grow with the sample size whenever the true model is sparse. Similar $L_2$ rate conditions can also be shown for non-linear models like Random Forests (\cite{Wager_Walther_2015}), Honest Random Forests (\cite{Wager_Athey_2018}) or forms of Deep Neural Nets (\cite{Farrell_Liang_Misra_2018}).
\begin{theorem}[Estimation and inference]\label{thm:estinf}
Suppose that $\psi(W,\eta)$ is a sum over the index $\tau$ of terms in the form 
\begin{align*}
\frac{q_1(X)}{q_1}\frac{G_{\tau}}{q_{G_\tau}(X)}\left(\tilde{Y}-m_{\tilde{Y}}(G_{\tau}=1,X)\right)+\psi^b(W,\eta)m_{\tilde{Y}}\left(G_{\tau}=1,X\right),
\end{align*}
$\mathbb{E}\left[\psi(W,\eta;\theta)\right]=0$ and Assumptions \ref{ass:estinfy}-\ref{ass:estinfjoint} hold. Then the cross-fitted estimator $\hat{\theta}$ obeys
\begin{align*}
\sqrt{N}\left(\hat{\theta}-\theta\right)=\frac{1}{\sqrt{N}}\sum_{i=1}^N\psi(W_i,\eta;\theta)+o_p(1)\overset{d}{\longrightarrow}\mathcal{N}(0,\sigma^2)
\end{align*}
where $\sigma^2=\mathbb{E}\left[\psi(W,\eta;\theta)^2\right]$.
\end{theorem}
\textit{Proof:} see Appendix \ref{app:estinf}.\\
Theorem \ref{thm:estinf} is our third main result. The cross-fitted estimator is asymptotically normal with influence function $\psi(W,\eta;\theta)$. Notice that the result implies that whenever we use the sample plug-in estimator of the efficient influence function the cross-fitted estimator asymptotically attains the low-dimensional variance lower bound if Assumptions \ref{ass:estinfy}-\ref{ass:estinfjoint} are satisfied. The following subsections contain several corollaries of Theorem \ref{thm:estinf} that summarize the convergence conditions and asymptotic behaviour of different cross-fitted semiparametric difference-in-differences estimators.
\subsection{Plug-in estimators}
Corollary \ref{cor:estcs} summarizes the implications of Theorem \ref{thm:estinf} for cross-sectional difference-in-differences estimators in settings (CS-1)-(CS-5) that use the efficient influence functions derived in Theorem \ref{thm:effcs}.
\begin{corollary}\label{cor:estcs}
Suppose that Assumptions \ref{ass:dgpcs}, \ref{ass:idcs}, \ref{ass:estinfy} and \ref{ass:estinfp} hold then
\begin{itemize}
\item[(a)] under setting (CS-1), $r=2$ and assuming that for all $d,t\in\{0,1\}$ $\epsilon_{p_{D=d,T=t}(X)}=o_p(1)$ and for $(d,t)\in\{(0,1),(1,0),(0,0)\}$ $\epsilon_{m_Y(d,t,X)}=o_p(1)$ and
\begin{align*}
\left(\epsilon_{p_{D=1,T=1}(X)}+\epsilon_{p_{D=d,T=t}(X)}\right)\times\epsilon_{m_Y(d,t,X)}=o_p\left(N^{-\frac{1}{2}}\right)
\end{align*}
\item[(b)] under the condition (CS-2) in Assumption \ref{ass:reldtxcs}, $r>2$ and assuming that $\epsilon_{p_D(X)}=o_p(1)$, $\epsilon_{p_T(X)}=o_p(1)$ and $\epsilon_{m_Y(d,t,X)}=o_p(1)$, $\left\lVert\hat{m}_Y(d,t,X)_{-k}-m_Y(d,t,X)\right\rVert_r=O_p(1)$ for all $d,t\in\{0,1\}$ and
\begin{align*}
&\epsilon_{p_D(X)}\times\epsilon_{p_T(X)}=o_p\left(N^{-\frac{1}{2}\frac{r}{r-1}}\right),\quad
\epsilon_{p_D(X)}\times\epsilon_{m_Y(0,1,X)}=o_p\left(N^{-\frac{1}{2}}\right),\\
&\epsilon_{p_T(X)}\times\epsilon_{m_Y(1,0,X)}=o_p\left(N^{-\frac{1}{2}}\right),\quad
\left(\epsilon_{p_D(X)}+\epsilon_{p_T(X)}\right)\times\epsilon_{m_Y(0,0,X)}=o_p\left(N^{-\frac{1}{2}}\right)
\end{align*}
\item[(c)] under the condition (CS-3) in Assumption \ref{ass:reldtxcs}, $r>2$ and assuming that $\epsilon_{p_D(1,X)}=o_p(1)$, $\epsilon_{p_D(0,X)}=o_p(1)$ and $\epsilon_{m_Y(d,t,X)}=o_p(1)$ for all $d,t\in\{0,1\}$ and
\begin{align*}
&\epsilon_{p_D(1,X)}\times\epsilon_{m_Y(0,1,X)}=o_p\left(N^{-\frac{1}{2}}\right),\quad
\left(\epsilon_{p_D(1,X)}+\epsilon_{p_D(0,X)}\right)\times\left(\epsilon_{m_Y(1,0,X)}+\epsilon_{m_Y(0,0,X)}\right)=o_p\left(N^{-\frac{1}{2}}\right)
\end{align*}
\item[(d)] under the condition (CS-4) in Assumption \ref{ass:reldtxcs}, $r=2$ and assuming that $\epsilon_{p_D(X)}=o_p(1)$ and $\epsilon_{m_Y(d,t,X)}=o_p(1)$ for all $d,t\in\{0,1\}$ and
\begin{align*}
\epsilon_{p_D(X)}\times\left(\epsilon_{m_Y(0,1,X)}+\epsilon_{m_Y(0,0,X)}\right)=o_p\left(N^{-\frac{1}{2}}\right)
\end{align*}
\item[(e)] under the condition (CS-5) in Assumption \ref{ass:reldtxcs}, $r=2$ and assuming that $\epsilon_{m_Y(d,t,X)}=o_p(1)$ for all $d,t\in\{0,1\}$
\end{itemize}
the sample plug-in estimators implied by $\mathbb{E}\left[\psi^*_{CS}(W;\theta)\right]=0$ are efficient estimators in the sense that they obey
\begin{align*}
\sqrt{N}\left(\hat{\theta}-\theta\right)=\frac{1}{\sqrt{N}}\sum_{i=1}^N\psi^*_{CS}(W_i;\theta)+o_p(1)\overset{d}{\longrightarrow}\mathcal{N}\left(0,\mathbb{E}\left[\psi^*_{CS}(W;\theta)^2\right]\right).
\end{align*}
\end{corollary}
\textit{Proof:} see Appendix \ref{app:estcs}.\\
The joint convergence conditions are more sophisticated for settings that are less restrictive with respect to the model assumptions on the relation between $D$, $T$ and $X$ imposed. For (CS-1) we need six, for (CS-2) and (CS-3) we need five, for (CS-4) we need two and for (CS-5) we need zero joint convergence rates to be satisfied. This hints at a trade-off between the robustness of the estimators with respect to the assumption on $D$, $T$ and $X$ and the robustness of the estimators with respect to first stage nuisance parameter convergence requirements. Further, we notice that for settings (CS-2) and (CS-3) higher-order moments $r>2$ have to exist. For setting (CS-3) this is a mere regularity condition. For setting (CS-2) it has some implications on the joint convergence condition of the two propensity scores. The more moments of the outcome exist, the closer the joint convergence condition at the usual $\sqrt{N}$ condition. This implies that for a bounded outcome the condition becomes $\epsilon_{p_D(X)}\times\epsilon_{p_T(X)}=o_p\left(N^{-\frac{1}{2}}\right)$.\footnote{For most outcomes in labour market applications the outcome is bounded. The condition might, however, become relevant for distributions with `fat tails' typically present in financial econometrics.} If the researcher is only willing to assume that a second moment of $Y$ exists, this implies that for both propensity scores parametric convergence rates are needed. We additionally notice that for the experimental setting (CS-5) the efficient estimator is not the simple difference in means estimator but a residualized version of it.\\
Corollary \ref{cor:estpa} summarizes the implications of Theorem \ref{thm:estinf} for panel difference-in-differences estimators in settings (PA-1) and (PA-2) that use the efficient influence functions derived in Theorem \ref{thm:effpa}.
\begin{corollary}\label{cor:estpa}
Suppose that Assumptions \ref{ass:dgppa}, \ref{ass:idpa}, \ref{ass:estinfy} and \ref{ass:estinfp} hold then
\begin{itemize}
\item[(a)] under the condition (PA-1), $r=2$ and assuming that $\epsilon_{p_{D}(X)}=o_p(1)$, $\epsilon_{m_{\Delta Y}(0,X)}=o_p(1)$ and
\begin{align*}
\epsilon_{p_{D}(X)}\times\epsilon_{m_{\Delta Y}(0,X)}=o_p\left(N^{-\frac{1}{2}}\right)
\end{align*}
\item[(b)] under the condition (PA-2) in Assumption \ref{ass:reldxpa}, $r=2$ and assuming that $\epsilon_{m_{\Delta Y}(d,X)}=o_p(1)$ for $d\in\{0,1\}$
\end{itemize}
the sample plug-in estimators implied by $\mathbb{E}\left[\psi^*_{PA}(W;\theta)\right]=0$ are efficient estimators in the sense that they obey
\begin{align*}
\sqrt{N}\left(\hat{\theta}-\theta\right)=\frac{1}{\sqrt{N}}\sum_{i=1}^N\psi^*_{PA}(W_i;\theta)+o_p(1)\overset{d}{\longrightarrow}\mathcal{N}\left(0,\mathbb{E}\left[\psi^*_{PA}(W;\theta)^2\right]\right).
\end{align*}
\end{corollary}
\textit{Proof:} see Appendix \ref{app:estpa}.\\
Compared to the cross-sectional case, the joint convergence condition are generally weaker for (PA-1) for two reasons. Firstly, the fact that we observe $\Delta Y$ only requires one projection on the difference of the outcomes instead of the difference between two projections on each single outcome. Secondly, the efficient influence function in Theorem \ref{thm:effpa} implies that $m_{\Delta Y}(1,X)$ is redundant. This allows us to obtain efficient difference-in-differences estimators for the panel under relatively weak conditions.
\subsection{Redundancy of nuisance parameters}
For the cross-sectional difference-in-differences estimators considered so far some nuisance parameters are redundant. For example in setting (CS-1) for all $x\in\mathcal{X}$ we have $\sum_{d=0}^1\sum_{t=0}^1p_{D=d,T=t}(x)=1$. In principle, we could therefore infer one of the four propensity scores from the other three. A well-known problem is that if one estimates the three propensity scores without requiring that the implied four propensity scores sum to one, Assumption \ref{ass:estinfp} might be violated because the fourth, implied propensity score is not guaranteed to be strictly greater than zero. A solution to this could be to explicitly require that the four propensity scores estimators sum to one for all $x\in\mathcal{X}$. Multinomial versions of the standard Logit regression or the Logit Lasso are available. To the knowledge of the author, solutions do not exist for more sophisticated machine-learning algorithms like Random Forests or Neural Nets. However, notice that $p_{D=d,T=t}(X)=p_{D=d}(t,X)p_{T=t}(X)=p_{T=t}(d,X)p_{D=d}(X)$ and that the propensity score estimators obey $\sum_{d=0}^1\sum_{t=0}^1\hat{p}_{D=d}(t,x)_{-k}\hat{p}_{T=t}(x)_{-k}=\sum_{d=0}^1\sum_{t=0}^1\hat{p}_{T=t}(d,x)_{-k}\hat{p}_{D=d}(x)_{-k}=1$ for all $x\in\mathcal{X}$ by construction. Corollary \ref{cor:redcs1} summarizes the properties of the two implied estimators.
\begin{corollary}\label{cor:redcs1}
Suppose that Assumptions \ref{ass:dgpcs}, \ref{ass:idcs}, \ref{ass:estinfy} and \ref{ass:estinfp} hold then under the condition (CS-1) and $r=2$, the estimators implied by the moment condition of the score functions
\begin{itemize}
\item[(a)]
\begin{align*}
\psi^{**}_{CS-1}(W,\eta;\theta)&=\frac{p_D(1,X)p_T(X)}{p_{DT}}\sum^1_{d=0}\sum^1_{t=0}(-1)^{(d+t)}\frac{G_{d,t}}{p_{D=d}(t,X)p_{T=t}(X)}\left(Y-m_Y(d,t,X)\right)\\
&+\frac{DT}{p_{DT}}\left(m_Y(X)-\theta\right)
\end{align*}
under the further conditions that $\epsilon_{p_D(1,X)}=o_p(1)$, $\epsilon_{p_D(0,X)}=o_p(1)$, $\epsilon_{p_T(X)}=o_p(1)$ and for all $(d,t)\in\{(0,1),(1,0),(0,0)\}$ $\epsilon_{m_Y(d,t,X)}=o_p(1)$ and
\begin{align*}
&\epsilon_{p_D(1,X)}\times\epsilon_{m_Y(0,1,X)}=o_p\left(N^{-\frac{1}{2}}\right)\quad\text{and}\\
&\left(\epsilon_{p_D(1,X)}+\epsilon_{p_D(0,X)}+\epsilon_{p_T(X)}\right)\times\left(\epsilon_{m_Y(1,0,X)}+\epsilon_{m_Y(0,0,X)}\right)=o_p\left(N^{-\frac{1}{2}}\right)
\end{align*}
\item[(b)]
\begin{align*}
\psi^{***}_{CS-1}(W,\eta;\theta)&=\frac{p_T(1,X)p_D(X)}{p_{DT}}\sum^1_{d=0}\sum^1_{t=0}(-1)^{(d+t)}\frac{G_{d,t}}{p_{T=t}(d,X)p_{D=d}(X)}\left(Y-m_Y(d,t,X)\right)\\
&+\frac{DT}{p_{DT}}\left(m_Y(X)-\theta\right)
\end{align*}
under the further conditions that $\epsilon_{p_T(1,X)}=o_p(1)$, $\epsilon_{p_T(0,X)}=o_p(1)$, $\epsilon_{p_D(X)}=o_p(1)$ and for all $(d,t)\in\{(0,1),(1,0),(0,0)\}$ $\epsilon_{m_Y(d,t,X)}=o_p(1)$ and
\begin{align*}
&\epsilon_{p_T(1,X)}\times\epsilon_{m_Y(1,0,X)}=o_p\left(N^{-\frac{1}{2}}\right)\quad\text{and}\\
&\left(\epsilon_{p_T(1,X)}+\epsilon_{p_T(0,X)}+\epsilon_{p_D(X)}\right)\times\left(\epsilon_{m_Y(0,1,X)}+\epsilon_{m_Y(0,0,X)}\right)=o_p\left(N^{-\frac{1}{2}}\right)
\end{align*}
\end{itemize}
obey
\begin{align*}
\sqrt{N}\left(\hat{\theta}-\theta\right)=\frac{1}{\sqrt{N}}\sum_{i=1}^N\psi^*_{CS-1}(W_i;\theta)+o_p(1)\overset{d}{\longrightarrow}\mathcal{N}\left(0,\mathbb{E}\left[\psi^*_{CS-1}(W;\theta)^2\right]\right).
\end{align*}
\end{corollary}
\textit{Proof:} The proof follows similarly to Corollary \ref{cor:estcs} (a).\\
Compared to Corollary \ref{cor:estcs} (a), the joint convergence conditions are increased from six to seven and the efficiency result is maintained. The scores rely on either $p_{D=d}(t,X)$ or $p_{T=t}(d,X)$. We therefore recommend to use $\psi^{**}_{CS-1}(W,\eta;\theta)$ when $\lvert p_T-0.5\rvert<\lvert p_D-0.5\rvert$ and $\psi^{***}_{CS-1}(W,\eta;\theta)$ otherwise.\\
The result of Corollary \ref{cor:estcs} (b)-(d) indicate that some of the outcome nuisances in settings (CS-2), (CS-3) and (CS-4) are redundant in the sense that they do not contribute to the joint convergence condition in Assumption \ref{ass:estinfjoint} (v). However, the outcome nuisances still need to converge in $L_2$ in order to satisfy Assumption \ref{ass:estinfjoint} (iv). The same applies for (CS-5) since the simple difference-in-means estimator represents an alternative estimator that does not rely on any first stage convergence conditions. We do not provide a separate result for case (CS-5). However, it is easy to see that the implied estimator without any outcome nuisances is just the difference-in-means estimator. Notice that as long as $Y$ is not independent of $X$ the difference-in-means estimator is \textit{not} an efficient estimator in a semiparametric sense. This has some practical relevance as in difference-in-differences the credibility of the design is often assessed by using placebo tests in some pre-periods without relying on covariates. Not rejecting the null might, however, just due to a higher than necessary standard error when using the difference-in-means estimator. The same applies for setting (PA-2). Since extensions for (CS-3) are also trivial, we focus on settings (CS-2) and (CS-4). Corollaries \ref{cor:redcs2} and \ref{cor:redcs4} summarize the implications when scores are used that do not contain some of the outcome nuisances. Generally, the implied cross-fitted sample plug-in estimators do not attain the variance lower bound but have otherwise desirable asymptotic properties under weaker conditions. This hints at another trade-off between the robustness of the estimator towards first stage convergence requirements and semiparametric efficiency.
\begin{corollary}\label{cor:redcs2}
Suppose that Assumptions \ref{ass:dgpcs}, \ref{ass:idcs}, \ref{ass:estinfy} and \ref{ass:estinfp} hold then under the condition (CS-2) in Assumption \ref{ass:reldtxcs}, $r=2$ and assuming that $\epsilon_{p_D(X)}=o_p(1)$, $\epsilon_{p_T(X)}=o_p(1)$ and $\epsilon_{m_Y(d,t,X)}=o_p(1)$ for all $(d,t)\in\{(0,1),(1,0),(0,0)\}$ and
\begin{align*}
&\epsilon_{p_D(X)}\times\epsilon_{m_Y(0,1,X)}=o_p\left(N^{-\frac{1}{2}}\right),\quad\epsilon_{p_T(X)}\times\epsilon_{m_Y(1,0,X)}=o_p\left(N^{-\frac{1}{2}}\right),\\
&\left(\epsilon_{p_D(X)}+\epsilon_{p_T(X)}\right)\times\epsilon_{m_Y(0,0,X)}=o_p\left(N^{-\frac{1}{2}}\right)
\end{align*}
the sample plug-in estimator using the score $\psi'_{CS-2}(W;\theta)=\frac{p_D(X)p_T(X)}{p_{DT}}\psi^{*a}_{CS-2}+\frac{DT}{p_{DT}}\left(m_Y(X)-\theta\right)$ obeys
\begin{align*}
\sqrt{N}\left(\hat{\theta}-\theta\right)=\frac{1}{\sqrt{N}}\sum_{i=1}^N\psi'_{CS-2}(W_i;\theta)+o_p(1)\overset{d}{\longrightarrow}\mathcal{N}\left(0,\mathbb{E}\left[\psi'_{CS-2}(W;\theta)^2\right]\right)
\end{align*}
with efficiency loss $\mathbb{E}\left[\psi'_{CS-2}(W;\theta)^2\right]-\mathbb{E}\left[\psi^{*}_{CS-2}(W;\theta)^2\right]=\Delta_{CS-1,CS-2}.$
\end{corollary}
\textit{Proof:} see Appendix \ref{app:redcs2}.\\
Corollary \ref{cor:redcs2} shows that for (CS-2) one may in principle get rid off the restrictive existence of higher-order moments requirement and the $L_2$ convergence condition for $m_Y(1,1,x)$ in Corollary \ref{cor:estcs} (b). However, this results in an efficiency loss. Notice that from Corollary \ref{cor:releffcs1} $\Delta_{CS-1,CS-2}$ represents the value of knowing that (CS-2) in Assumption \ref{ass:reldtxcs} is true relative to not making any assumptions. Hence, when using score $\psi'_{CS-2}(W;\theta)$ all efficiency gains from the stronger setting (CS-2) relative to (CS-1) are exchanged for relatively weak first stage convergence conditions.
\begin{corollary}\label{cor:redcs4}
Suppose that Assumptions \ref{ass:dgpcs}, \ref{ass:idcs}, \ref{ass:estinfy} and \ref{ass:estinfp} hold then under the condition (CS-4) in Assumption \ref{ass:reldtxcs}, $r=2$ and assuming that $\epsilon_{p_D(X)}=o_p(1)$ and $\epsilon_{m_Y(d,t,X)}=o_p(1)$ for all $(d,t)\in\{(0,1),(0,0)\}$ and
\begin{align*}
\epsilon_{p_D(X)}\times\left(\epsilon_{m_Y(0,1,X)}+\epsilon_{m_Y(0,0,X)}\right)=o_p\left(N^{-\frac{1}{2}}\right)
\end{align*}
the sample plug-in estimator using the score 
\begin{align*}
\psi'_{CS-4}(W;\theta)=\psi^*_{CS-4}(W;\theta)+\frac{D}{p_D}\left(\frac{T}{p_T}-1\right)m_Y(1,1,X)-\frac{D}{p_D}\left(\frac{1-T}{1-p_T}-1\right)m_Y(1,0,X)
\end{align*}
obeys
\begin{align*}
\sqrt{N}\left(\hat{\theta}-\theta\right)=\frac{1}{\sqrt{N}}\sum_{i=1}^N\psi'_{CS-4}(W_i;\theta)+o_p(1)\overset{d}{\longrightarrow}\mathcal{N}\left(0,\mathbb{E}\left[\psi'_{CS-4}(W;\theta)^2\right]\right)
\end{align*}
with efficiency loss $\mathbb{E}\left[\psi'_{CS-4}(W;\theta)^2\right]-\mathbb{E}\left[\psi^{*}_{CS-4}(W;\theta)^2\right]=\mathbb{E}\left[\frac{p_D(X)^2}{p_D^2}\frac{\left(\sqrt{\frac{1-p_T}{p_T}}m_Y(1,1,X)+\sqrt{\frac{p_T}{1-p_T}}m_Y(1,0,X)\right)^2}{p_D(X)}\right].$
\end{corollary}
\textit{Proof:} see Appendix \ref{app:redcs4}.\\
Corollary \ref{cor:redcs4} provides the convergence conditions when we use a score where $m_Y(1,1,x)$ and $m_Y(1,0,x)$ are redundant. This results in convergence conditions similar to those of the score $\psi^{*}_{PA-1}(W;\theta)$. However, also the efficiency loss can be relatively huge.
\section{Application}\label{sec:app}
To illustrate the practical relevance of the proposed method, we revisit \textcite{Angrist_Acemoglu_2001}. The paper is concerned with the theoretically ambiguous effect of increased employment protection for disabled workers on weeks worked (for more details see the paper). An empirical evaluation of the Americans with Disabilities Act reform introduced in 1991 is used to test the theory using data from the Current Population Survey (CPS), a repeated cross-section.\\
As in the paper, we define $D$ as being disabled. $Y$ are the weeks worked in the respective year. The outcome is therefore bounded between 0 and 52. In the original paper several post-reform years are considered. Since the credibility of the common trend assumption might be questionable for years well after the reform, we focus on 1992 as the post-reform period ($T=1$). As in the original paper we use the years 1988-1990 for $T=0$. All CPS data is retrieved from Joshua Angrist's data archive\footnote{\href{https://economics.mit.edu/faculty/angrist/data1/data/aceang01}{https://economics.mit.edu/faculty/angrist/data1/data/aceang01}}. We consider three different sets of covariates. An overview of the different variable sets is provided in Table \ref{tab:spec}.
\begin{table}[h!]
\centering
\caption{Covariate specifications}
\label{tab:spec}
\begin{threeparttable}
\begin{tabular}{lp{10cm}l}
  \toprule
specification & covariates used & \# of covariates\\ 
  \midrule
 original & sex, age, race group, education group, region & 14 \\
 baseline & sex, age, race group, education group, marital status, class of worker, major industry, major occupation, state, central city MSA status & 108\\
 extended & sex, age, race group, education group, marital status, class of worker, major industry, major occupation, state, central city MSA status, longest job class of worker, longest job major occupation, longest job major industry, number of employers, unemployment compensation benefit value, supplemental security income amount received, public assistance or welfare value received, social security payments received, veteran status, veterans payment income, survivor's income received, value of other income, value of workers' compensation for job related illness or injury, retirement income, health insurance group, medicare coverage, medicaid coverage, coverage by military health care & 161\\
\bottomrule \bottomrule
\end{tabular}
\end{threeparttable}
\end{table}
The variable set labelled as `original' is constructed using the covariates that are also included in the specifications of \textcite{Angrist_Acemoglu_2001}. The variable sets `baseline' and `extended' use some further covariates available from the CPS. The `baseline' specification disaggregates the region variable used in the original dataset to control for geographically different common trends. Instead, we include state dummies and dummies indicating whether the individual lives in a metropolitain area. Additional controls on marital status, class of worker, industry and occupation should help to control for non-parallel trends between disabled and non-disabled. For example, if people who report a disability self-select more often into a certain sector or industry then any underlying structural change in this sector needs to be controlled for in order to guarantee the common trend assumption to hold. This might be a point that is of more general interest in empirical economics. In labour market applications the common trend is often valid only after conditioning on variables that are available at different aggregation levels e.g. geographic or sector dummies. The `extended' specification includes further covariates on employment history, social welfare payments and health insurance status.
\begin{table}[b!]
\centering
\caption{Subsample sizes}
\label{tab:sub}
\begin{threeparttable}
\begin{tabular}{llllllllll}
  \toprule
$N(0)$ & $N(1)$ & $p_D$ & $p_T$ & $p_D(1)$ & $p_D(0)$ & $p_{D=1,T=1}$ & $p_{D=0,T=1}$ & $p_{D=1,T=0}$ & $p_{D=0,T=0}$\\ 
  \midrule
206058 & 70069 & 6.19 \% & 25.38 \% & 6.49 \% & 6.08 \% & 1.65 \% & 23.73 \% & 4.54 \% & 70.08 \% \\
\bottomrule \bottomrule
\end{tabular}
\end{threeparttable}
\end{table}
Table \ref{tab:sub} shows that the merged sample consists of 276127 observations (for all specifications) and is highly imbalanced. We observe that the share of $D$ varies between $T=0$ and $T=1$. Given the large sample size, this indicates that assumptions (CS-4) and (CS-5) are likely to be violated.
\subsection{Estimators considered}
We consider different estimators using the different score functions outlined in Section \ref{sec:estinf} with different first stage estimators.\\
Our preferred estimator is an Ensemble Learner that weights Lasso and Random Forest predictions by using out-of-sample MSE optimal weights. For the Lasso we allow for polynomials up to order four and all two way interactions. The Random Forest is an ensemble of  regression trees and therefore implicitly contains higher-order terms. For both estimators we use the default settings in the \textit{glmnet} and \textit{ranger} \R{}-packages. Using an ensemble of machine-learning estimators has several important advantages. Firstly, Lasso and Random Forests are designed for different DGPs. Whereas the Lasso allows for some form of smoothing, we expect a tree-based estimator to work well with strong non-linearities. Secondly, the Ensemble gives more weight to the single predictor that works best and therefore should be less dependent on the particular tuning parameter choices of the Lasso and the Random Forest. Thirdly, the strong imbalances in our sample indicate that using single propensity score estimators might result in extreme weights. By combining two predictors the likelihood of generating extreme weights is reduced.\\
We compare the Ensemble Learner to the single predictors Lasso and Random Forest. Due to computational constraints, we restrict this analysis to the `baseline' set of covariates. Further, we compare the Ensemble to parametric models for the propensity scores and the outcome nuisances. For the propensity scores we use Logit and for the outcome nuisances standard linear regression. We include covariates in levels. As a benchmark, we also consider \posscite{Abadie_2005} Inverse Probability Weighting (IPW) difference-in-differences estimator with Logit regression for the propensity score. 
\subsection{Placebo tests}
As usual in the difference-in-differences literature, we run some placebo experiments for period 1988/89.
\begin{table}[h!]
\centering
\caption{Placebo tests 1988/89 under (CS-5)}
\label{tab:plcebocs5}
\begin{threeparttable}
\begin{tabular}{lllll}
 & $\eta$ estimator & original & baseline & extended\\ 
  \cmidrule(lr){2-2} \cmidrule(lr){3-5}
$\psi^*_{CS-5}(W,\eta;\theta)$ & Ensemble & -1.410 & -0.956 & -0.600 \\
 & & (0.464) & (0.344) & (0.218)\\
$\psi^{*}_{CS-5}(W,\eta;\theta)$ & Linear & -1.141 & -1.077 & -1.257\\
 & & (0.551) & (0.506) & (0.690) \\
Mean differences & & -0.801 & -0.801 & -0.801 \\
 & & (0.652) & (0.652) & (0.652) \\
\bottomrule \bottomrule
\end{tabular}
\begin{tablenotes}
\item Results for estimators with the Ensemble were obtained using the cross-fitting procedure in Section \ref{sec:estinf} with $K=2$. The Ensemble Learner comprises Lasso and Random Forest. For Lasso the penalty term was chosen such that the cross-validation criterion was minimized. The Ensemble weights were chosen by minimizing out-of-sample MSE. Results for estimators with Linear regression were obtained by using the sample plug-in estimator without cross-fitting. The sample for the placebo tests contains $N=135174$ observations. Standard errors are in parenthesis.
\end{tablenotes}
\end{threeparttable}
\end{table}
Table \ref{tab:plcebocs5} shows the results under (CS-5). Notice that the simple differences-in-means estimator does not hint at a violation of the specification that does not account for any imbalances. This is not because the effect is economically small but due to the comparatively high standard error. If the efficient score $\psi^*_{CS-5}(W,\eta;\theta)$ is used, the point estimators are in the same range but -- in line with theory in Section \ref{sec:ideff} -- the standard errors are substantially decreased leading to significant results. This shows that using the efficient score function rather than the simple mean differences estimator may lead to opposite conclusions regarding the credibility of the design.\footnote{We notice that the asymptotic standard error might, however, be a more accurate approximation of the finite sample standard error in case of the difference-in-means estimator.} The result also highlights the importance of discriminating the two roles covariates can have in semiparametric difference-in-differences estimation. First of all, they may be included to improve the reliability of the Conditional Common Trends (see Assumptions \ref{ass:idcs} and \ref{ass:idpa}). Second of all, under some assumptions, they should be included to improve on the efficiency of the derived estimator. Further, we notice that the standard error for the Ensemble Learner decreases when more covariates are added to the model, whereas the standard error when using the linear model increases for the `extended' specification. The standard errors are also generally larger. This indicates that the Ensemble Learner is more effective in predicting the conditional outcomes leading to a stronger variance reduction of the residualized estimator.\\
From these results we conclude that incorporating covariates seems to be necessary in our example. Notice that due to the strong imbalances in our application, especially propensity scores for $D=1$, $T=1$ are hard to predict. Results of placebo tests for our estimators are reported in Table \ref{tab:placebocs}. For estimators that rely on the Ensemble Learner none of the different specifications used hint at a violation of the conditional common trend assumption. Moreover, the standard errors are reduced when a higher number of covariates is included, indicating that the Ensemble is effective in extracting the additional information. When covariates in `baseline' and `extended' are used, estimators under (CS-2) and (CS-4) show substantially decreased standard errors compared to estimators that rely on (CS-1). This is in line with the theoretical argument in Corollary \ref{cor:releffcs1}. When using the same scores with the same set of variables most estimators that rely on parametric models hint at a potential violation of the conditional common trend assumption. With the exception of IPW, estimators implied by scores that only rely on the propensity scores $p_D(X)$ and $p_T(X)$ ((CS-2), (CS-4)) generally do not indicate a violation of the conditional common trend assumption. The same can be observed for the single nuisance predictors. We conclude that scores that rely on sophisticated convergence conditions are less robust to potential violations of these conditions in practice. However, it seems that the efficiency-estimation trade-off seems to be less of a concern when high-quality predictors like the Ensemble are used. Moreover, even when using a parametric model and a large number of covariates the double robust scores give results whereas IPW explodes for the `extended' specification.
\begin{table}[p]
\centering
\caption{Placebo tests 1988/89 under (CS-1)-(CS-4)}
\label{tab:placebocs}
\begin{threeparttable}
\begin{tabular}{lllllllll}
 & \multicolumn{3}{c}{Ensemble} & \multicolumn{3}{c}{Linear/Logit regression} & Forest & Lasso \\
 \cmidrule(lr){2-4} \cmidrule(lr){5-7} \cmidrule(lr){8-8} \cmidrule(lr){9-9} 
 & original & baseline & extended & original & baseline & extended & baseline & baseline \\
 \cmidrule(lr){2-4} \cmidrule(lr){5-7} \cmidrule(lr){8-8} \cmidrule(lr){9-9} 
$\psi^*_{CS-1}(W,\eta;\theta)$ & -0.368 & 0.223 & -0.167 & -1.338 & -3.875 & -4.832 & 7.564 & -10.850 \\
 & (0.487) & (0.496) & (0.457) & (0.546) & (0.435) & (0.598) & (0.857) & (0.410) \\
$\psi^{**}_{CS-1}(W,\eta;\theta)$ & -0.359 & 0.255 & -0.527 & -1.425 & -3.598 & -8.640 & 24.452 & -15.053\\
 & (0.487) & (0.470) & (0.367) & (0.544) & (0.443) & (0.516) & (1.363) & (0.365) \\
$\psi^{***}_{CS-1}(W,\eta;\theta)$ & -0.390 & 0.084 & -0.352 & -1.296 & -4.033 & -6.802 & 4.205 & -0.753 \\
 & (0.481) & (0.385) & (0.287) & (0.548) & (0.431) & (0.554) & (0.765) & (0.618)\\
$\psi^{*}_{CS-2}(W,\eta;\theta)$ & -0.470 & -0.022 & -0.391 & -0.977 & -0.994 & 0.849 & 9.365 & -0.772\\
 & (0.477) & (0.379) & (0.294) & (0.559) & (0.510) & (0.733) & (0.875) & (0.605)\\
$\psi^{'}_{CS-2}(W,\eta;\theta)$ & -0.429 & 0.008 & -0.406 & -0.992 & -1.070 & 0.870 & 11.121 & -0.819\\
 & (0.477) & (0.381) & (0.296) & (0.556) & (0.524) & (0.730) & (0.961) & (0.616)\\
$\psi^{*}_{CS-3}(W,\eta;\theta)$ & -0.317 & 0.271 & -0.484 & -1.847 & -3.817 & -10.331 & 7.275 & -14.973\\
 & (0.487) & (0.474) & (0.371) & (0.528) & (0.433) & (0.481) & (0.796) & (0.364)\\
$\psi^{*}_{CS-4}(W,\eta;\theta)$ & -0.685 & -0.378 & -0.312 & -1.138 & -1.271 & -0.749 & 0.363 & -0.762\\
 & (0.477) & (0.385) & (0.277) & (0.542) & (0.494) & (0.680) & (0.594) & (0.586)\\
$\psi^{'}_{CS-4}(W,\eta;\theta)$ & -0.645 & -0.416 & -0.450 & -0.832 & -1.027 & -0.328 & 0.719 & -0.413\\
 & (0.620) & (0.631) & (0.627) & (0.604) & (0.602) & (0.602) & (0.603) & (0.603)\\
 IPW & & & & -1.808 & -0.558 & - & & \\
 & & & & (0.686) & (0.658) & - & & \\
\bottomrule \bottomrule
\end{tabular}
\begin{tablenotes}
\item Results for estimators with the Ensemble, the Random Forest and the Lasso were obtained using the cross-fitting procedure in Section \ref{sec:estinf} with $K=2$. The Ensemble Learner comprises Lasso and Random Forest.
For Lasso the penalty term was chosen such that the cross-validation criterion was minimized. The Ensemble weights were
chosen by minimizing out-of-sample MSE. Results for estimators with Linear/Logit regression were obtained by using the sample plug-in estimator without cross-fitting. The Logit was used for the propensity scores and linear regression for the outcome nuisances. The sample for the placebo tests contains $N=135174$ observations. Standard errors are in parenthesis.
\end{tablenotes}
\end{threeparttable}
\end{table}
\begin{table}[p]
\centering
\caption{Results for (CS-1)-(CS-4)}
\label{tab:resultscs}
\begin{threeparttable}
\begin{tabular}{lllllllll}
 & \multicolumn{3}{c}{Ensemble} & \multicolumn{3}{c}{Linear/Logit regression} & Forest & Lasso \\
 \cmidrule(lr){2-4} \cmidrule(lr){5-7} \cmidrule(lr){8-8} \cmidrule(lr){9-9} 
 & original & baseline & extended & original & baseline & extended & baseline & baseline \\
 \cmidrule(lr){2-4} \cmidrule(lr){5-7} \cmidrule(lr){8-8} \cmidrule(lr){9-9} 
$\psi^*_{CS-1}(W,\eta;\theta)$ & -0.904 & -0.855 & 0.031 & 1.049 & 2.118 & 13.959 & 13.123 & 17.558 \\
 & (0.376) & (0.351) & (0.273) & (0.458) & (0.465) & (0.782) & (0.691) & (0.795) \\
$\psi^{**}_{CS-1}(W,\eta;\theta)$ & -0.828 & -0.689 & 0.017 & 0.785 & 0.928 & 10.466 & 14.793 & 11.546\\
 & (0.377) & (0.351) & (0.209) & (0.453) & (0.442) & (0.720) & (0.720) & (0.680)\\
$\psi^{***}_{CS-1}(W,\eta;\theta)$ & -0.837 & -0.356 & -0.096 & 1.144 & 0.978 & 6.457 & 23.419 & -0.120\\
 & (0.375) & (0.318) & (0.184) & (0.459) & (0.443) & (0.650) & (0.887) & (0.478)\\
$\psi^{*}_{CS-2}(W,\eta;\theta)$ & -0.856 & -0.422 & -0.169 & -0.456 & -0.845 & -1.746 & -11.963 & -0.987\\
 & (0.365) & (0.307) & (0.183) & (0.434) & (0.429) & (0.542) & (0.432) & (0.491)\\
$\psi^{'}_{CS-2}(W,\eta;\theta)$ & -0.792 & -0.407 & -0.123 & -0.453 & -0.819 & -1.695 & -7.549 & -0.895\\
 & (0.364) & (0.309) & (0.185) & (0.432) & (0.411) & (0.517) & (0.371) & (0.466)\\
$\psi^{*}_{CS-3}(W,\eta;\theta)$ & -0.934 & -0.748 & -0.061 & -0.121 & 0.135 & 8.760 & 34.956 & 10.029\\
 & (0.378) & (0.345) & (0.198) & (0.434) & (0.426) & (0.688) & (1.127) & (0.648)\\
$\psi^{*}_{CS-4}(W,\eta;\theta)$ & -0.869 & -0.294 & -0.118 & -1.013 & -1.334 & -2.409 & -1.385 & -1.465\\
 & (0.384) & (0.322) & (0.185) & (0.440) & (0.436) & (0.553) & (0.494) & (0.502)\\
$\psi^{'}_{CS-4}(W,\eta;\theta)$ & 0.903 & 0.800 & 0.284 & 0.834 & 0.501 & 0.142 & 0.842 & 0.793\\
 & (0.498) & (0.514) & (0.490) & (0.486) & (0.485) & (0.485) & (0.485) & (0.487)\\
  IPW & & & & 1.470 & 0.148 & - & & \\
 & & & & (0.546) & (0.529) & - & & \\
\bottomrule \bottomrule
\end{tabular}
\begin{tablenotes}
\item Results for estimators with the Ensemble, the Random Forest and the Lasso were obtained using the cross-fitting procedure in Section \ref{sec:estinf} with $K=2$. The Ensemble Learner comprises Lasso and Random Forest.
For Lasso the penalty term was chosen such that the cross-validation criterion was minimized. The Ensemble weights were
chosen by minimizing out-of-sample MSE. Results for estimators with Linear/Logit regression were obtained by using the sample plug-in estimator without cross-fitting. The Logit was used for the propensity scores and linear regression for the outcome nuisances. The sample contains $N=276127$ observations. Standard errors are in parenthesis.
\end{tablenotes}
\end{threeparttable}
\end{table}
\subsection{Main results}
The placebo tests for the different estimators suggest some trustworthiness of estimators that use the Ensemble Learner for first stage prediction. Table \ref{tab:resultscs} summarizes the main results for the difference-in-differences estimators considered. For some of the specifications for the single predictors and the linear estimators we obtain extreme results. In general the same conclusions as for the placebo test apply. We notice that the particular first stage estimator and the estimation-robustness of the score used can drastically shift the results. Whereas estimators for (CS-1)-(CS-3) that use the Ensemble Learner unambiguously give negative or insignificant effects of reasonable size, again some of the other estimators explode. In addition, the Ensemble Learner based estimators mostly become insignificant when more covariates are included in the model. In contrast, to estimators with parametric first stages, this is not due to an increase in standard errors if more covariates are included in the model. Rather, the decreased standard errors indicate that the Ensemble Learner allows to effectively exhaust the information of growing covariate sets. Lastly, we notice that for specifications `baseline' and `extended' the efficiency-robustness trade-off from Corollary \ref{cor:releffcs1} becomes relevant. Scores under setting (CS-1) generally have higher standard errors compared to efficient scores in settings (CS-2), (CS-3) and (CS-4). While for (CS-4) using $\psi^{'}_{CS-4}(W,\eta;\theta)$ leads to a substantially higher standard error compared to the efficient score, for (CS-2) there is barely a difference. Also the point estimators are similar. This is expected since the score $\psi^{'}_{CS-2}(W,\eta;\theta)$ converges under similar conditions when the outcome is bounded.
\section{Conclusion}\label{sec:conc}
Semiparametric difference-in-differences estimation is a non-trivial endeavour. In this study we highlight the importance of different assumptions in semiparametric difference-in-differences models. Our results show that efficiency bounds may strongly depend on the model assumptions imposed and the data that is available. In particular, we show that there is a trade-off between the strength of the assumptions imposed and the variance lower bound that can be achieved. For estimation we provide easy-to-check conditions to derive the required convergence rates for a broad class of estimation problems. Our theoretical results allow to integrate scores with sophisticated adjustment terms in the double machine-learning framework. Further, we show that the different semiparametric models imply estimators with different properties. Estimators that are more robust against the model assumptions imposed also rely on more sophisticated conditions for first stage prediction. Some of these conditions can be relaxed when we give up on asymptotically attaining the efficiency bound. An empirical example shows that our proposed estimators are useful in practice. However, estimation results might be highly sensitive regarding the choice of the first stage nuisance parameter predictor. Placebo tests indicate that, in contrast to other choices, our proposed Ensemble Learner performs well.\\
Some interesting problems are beyond the scope of this study and have to be left for further research. The performance of the estimators proposed might depend on the parameter $K$. Some theoretical results how to optimally choose this parameter would be helpful. Our theoretical results for the different estimators suggest that the finite sample performance of the point estimators and the coverage probabilities of the variance estimators might be relatively diverse and depend on the particular DGP considered. Monte Carlo simulations could shed some more light on this subject. Also this study is limited to two time periods and two groups. Some recent advances in the semiparametric difference-in-differences literature (e.g., \cite{Callaway_SantAnna_2018}, \cite{GoodmanBacon_2018}) comprise extensions to more complicated adoption patterns. Further, in practice panel and cross-sectional data are often combined (e.g. in rotating panels). The results in Section \ref{sec:ideff} suggest that for these kind of data combination problems efficiency gains are possible if neither the panel structure is neglected nor the cross-sectional data is thrown away. A generalization of the efficiency theory provided in this study to these settings represents yet another avenue for further research.
\printbibliography
\begin{appendix}
\section{Proofs for Section \ref{sec:ideff}}
\subsection{Proof of Theorem \ref{thm:effcs}}\label{app:effcs}
We observe the data $W=(Y,D,T,X)$. For the joint distribution of the data consider a regular parametric submodel indexed by $\beta$. The density under the submodel can then be written as
\begin{align*}
f_W(w;\beta)=f_{Y|D,T,X}(y|d,t,x;\beta)f_{D,T|X}(d,t|x;\beta)f_{X}(x;\beta)
\end{align*}
which equals $f_W(w)$ at $\beta=\beta_0$.\\
The score function is defined as $S(y,d,t,x;\beta_0)=\frac{\partial \log f_W(w;\beta_0)}{\partial \beta}$ and we obtain $S(y,d,t,x;\beta_0)=S_y(y,d,t,x;\beta_0)+S_{d,t}(d,t,x;\beta_0)+S_x(x;\beta_0)$ with
\begin{align*}
S&_y(y,d,t,x;\beta_0)=\frac{\partial \log f_{Y|D,T,X}(y|d,t,x;\beta_0)}{\partial \beta}\\
S&_{d,t}(d,t,x;\beta_0)=\frac{\partial \log f_{D,T|X}(d,t|x;\beta_0)}{\partial \beta}\\
S&_x(x;\beta_0)=\frac{\partial \log f_{X}(x;\beta_0)}{\partial \beta}
\end{align*}
where 
\begin{align*}
S_y(y,d,t,x;\beta_0)&=\sum_{d=0}^1\sum_{t=0}^1g_{d,t}S_y(d,t,x;\beta_0)
\end{align*}
and $S_{d,t}(d,t,x;\beta_0)$ depends on the settings (CS-1) to (CS-5) on the relation between $D$, $T$ and $X$.\\~\\
For all regular parametric submodels the variance lower bound for a model is the second moment of the projection of a function $\psi^{*}_{CS}(W;\theta)$ (with $\mathbb{E}[\psi^{*}_{CS}(W;\theta)]=0$ and an existing second moment) on the tangent space $\mathcal{T}$ that satisfies
\begin{align*}
\frac{\partial \theta(\beta_0)}{\partial \beta}=\mathbb{E}\left[\psi^{*}_{CS}(W;\theta)S(Y,D,T,X;\beta_0)\right].
\end{align*}
When $\psi^{*}_{CS}(W;\theta)\in\mathcal{T}$, the projection on $\mathcal{T}$ is the function itself and therefore the variance lower bound for the model is given by $\mathbb{E}[\psi^{*}_{CS}(W;\theta)^2]$.
\subsubsection*{Proof for CS-1}
Under the conditions in (CS-1)
\begin{align*}
S_{d,t}(d,t,x;\beta_0)=\sum_{d=0}^1\sum_{t=0}^1\frac{g_{d,t}}{p_{D=d,T=t}(x)}\dot{p}_{D=d,T=t}(x;\beta_0).
\end{align*}
The tangent space of the model is characterized by the set of functions that are mean zero and satisfy the structure of the score function
\begin{align*}
\mathcal{T}&=\Bigg\{\sum_{d=0}^1\sum_{t=0}^1\left(g_{d,t}S_y(d,t,x)+\frac{g_{d,t}}{p_{D=d,T=t}(x)}\dot{p}_{D=d,T=t}(x)\right)+S_x(x)\Bigg\}
\end{align*}
for any functions $S_y(d,t,x)$, $\dot{p}_{D=d,T=t}(x)$ and $S_x(x)$ that satisfy
\begin{align*}
\mathbb{E}&\left[G_{d,t}S_y(D,T,X)\right]=\mathbb{E}\left[p_{D=d,T=t}(X)\mathbb{E}\left[S_y(d,t,X)|D=d,T=t,X\right]\right]=0\\
\mathbb{E}&\left[\sum_{d=0}^1\sum_{t=0}^1\frac{G_{d,t}}{p_{D=d,T=t}(X)}\dot{p}_{D=d,T=t}(X)\right]=\mathbb{E}\left[\sum_{d=0}^1\sum_{t=0}^1\dot{p}_{D=d,T=t}(X)\right]=0\\
\mathbb{E}&\left[S_x(X)\right]=0
\end{align*}
where the first and last equality follow by the mean zero property of the score function and the second equality by the fact that $\sum_{d=0}^1\sum_{t=0}^1\dot{p}_{D=d,T=t}(x)=0$ since $\sum_{d=0}^1\sum_{t=0}^1p_{D=d,T=t}(x)=1$.\\
The parameter $\theta$ is pathwise differentiable. For the parametric submodel we have
\begin{align*}
\theta(\beta)=\frac{\sum_{d=0}^1\sum_{t=0}^1(-1)^{(d+t)}\int\int y f_{Y|D,T,X}(y|d,t,x;\beta)p_{D=1,T=1}(x;\beta)f_X(x;\beta)dydx}{\int p_{D=1,T=1}(x;\beta)f_X(x;\beta)dx}
\end{align*}
and at $\beta=\beta_0$
\begin{align*}
\frac{\partial \theta(\beta_0)}{\partial \beta}&=\frac{1}{p_{DT}}\left(\sum_{d=0}^1\sum_{t=0}^1(-1)^{(d+t)}\int\int yS_y(d,t,x;\beta_0)f_{Y|D,T,X}(y|d,t,x)p_{D=1,T=1}(x)f_X(x)dydx\right)\\
&+\frac{1}{p_{DT}}\int(m_Y(X)-\theta)\left(\dot{p}_{D=1,T=1}(x;\beta_0)+p_{D=1,T=1}(x)S_x(x;\beta_0)\right)f_X(x)dx.
\end{align*}
We consider the function
\begin{align*}
\psi^{*}_{CS-1}(W;\theta)=\frac{p_{D=1,T=1}(X)}{p_{DT}}\left(\sum_{d=0}^1\sum_{t=0}^1(-1)^{(d+t)}\frac{G_{d,t}}{p_{D=d,T=t}(X)}(Y-m_Y(d,t,X))\right)+\frac{DT}{p_{DT}}(m_Y(X)-\theta).
\end{align*}
Notice that $\psi^{*}_{CS-1}(W;\theta)\in\mathcal{T}$. Also for any $D=d$, $T=t$ we obtain
\begin{align*}
\mathbb{E}&\left[\frac{p_{D=1,T=1}(X)}{p_{DT}}\frac{G_{d,t}}{p_{D=d,T=t}(X)}(Y-m_Y(d,t,X))\times S(Y,D,T,X;\beta_0)\right]\\
&=\mathbb{E}\left[\frac{p_{D=1,T=1}(X)}{p_{DT}}\mathbb{E}\left[YS_y(d,t,X;\beta_0)|D=d,T=t,X\right]\right]\\
&=\frac{1}{p_{DT}}\int\int yS_y(d,t,x;\beta_0)f_{Y|D,T,X}(y|d,t,x)p_{D=1,T=1}(x)f_X(x)dydx
\end{align*}
which follows from the fact that $\mathbb{E}\left[S_y(d,t,X;\beta_0)|D=d,T=t,X\right]=0$. Further,
\begin{align*}
\mathbb{E}&\left[\frac{DT}{p_{DT}}(m_Y(X)-\theta)\times S(Y,D,T,X;\beta_0)\right]\\
&=\frac{1}{p_{DT}}\mathbb{E}\left[(m_Y(X)-\theta)\left(\dot{p}_{D=1,T=1}(X;\beta_0)+p_{D=1,T=1}(X)S_x(X;\beta_0)\right)\right]\\
&=\frac{1}{p_{DT}}\int (m_Y(X)-\theta)\left(\dot{p}_{D=1,T=1}(x;\beta_0)+p_{D=1,T=1}(x)S_x(x;\beta_0)\right)f_X(x)dx.
\end{align*}
It follows that $\psi^{*}_{CS-1}(W;\theta)$ is the efficient influence function and the variance lower bound for (CS-1) is $\mathbb{E}[\psi^{*}_{CS-1}(W;\theta)^2]$.
\subsubsection*{Proof for CS-2}
Under the conditions in (CS-2)
\begin{align*}
S_{d,t}(d,t,x;\beta_0)=\left(\frac{d}{p_D(x)}-\frac{1-d}{1-p_D(x)}\right)\dot{p}_D(x;\beta_0)+\left(\frac{t}{p_T(x)}-\frac{1-t}{1-p_T(x)}\right)\dot{p}_T(x;\beta_0).
\end{align*}
The tangent space of the model is characterized by the set of functions that are mean zero and satisfy the structure of the score function
\begin{align*}
\mathcal{T}&=\Bigg\{\sum_{d=0}^1\sum_{t=0}^1(g_{d,t}S_y(d,t,x))+(d-p_D(x))a(x)+(t-p_T(x))b(x)+S_x(x)\Bigg\}
\end{align*}
for any functions $S_y(d,t,x)$ and $S_x(x)$ that satisfy
\begin{align*}
\mathbb{E}&\left[S_y(d,t,X)|D=d,T=t,X\right]=0\\
\mathbb{E}&\left[S_x(X)\right]=0
\end{align*}
and any square integrable functions $a(x)$ and $b(x)$.\\
The parameter $\theta$ is pathwise differentiable. For the parametric submodel we have
\begin{align*}
\theta(\beta)=\frac{\sum_{d=0}^1\sum_{t=0}^1(-1)^{(d+t)}\int\int y f_{Y|D,T,X}(y|d,t,x;\beta)p_{D}(x;\beta)p_{T}(x;\beta)f_X(x;\beta)dydx}{\int p_{D}(x;\beta)p_{T}(x;\beta)f_X(x;\beta)dx}
\end{align*}
and at $\beta=\beta_0$
\begin{align*}
\frac{\partial \theta(\beta_0)}{\partial \beta}&=\frac{1}{p_{DT}}\left(\sum_{d=0}^1\sum_{t=0}^1(-1)^{(d+t)}\int\int yS_y(d,t,x;\beta_0)f_{Y|D,T,X}(y|d,t,x)p_{D}(x)p_T(x)f_X(x)dydx\right)\\
&+\frac{1}{p_{DT}}\int(m_Y(X)-\theta)\left(\dot{p}_{D}(x;\beta_0)p_T(x)+p_{D}(x)\dot{p}_T(x;\beta_0)+p_D(x)p_T(x)S_x(x;\beta_0)\right)f_X(x)dx.
\end{align*}
We consider the function
\begin{align*}
\psi^{*}_{CS-2}(W;\theta)&=\frac{p_{D}(X)p_T(X)}{p_{DT}}\left(\sum_{d=0}^1\sum_{t=0}^1(-1)^{(d+t)}\frac{G_{d,t}}{p_{D=d}(X)p_{T=t}(X)}(Y-m_Y(d,t,X))\right)\\
&+\left(\frac{p_T(X)(D-p_D(X))}{p_{DT}}+\frac{p_D(X)(T-p_T(X))}{p_{DT}}+\frac{p_D(X)p_T(X)}{p_{DT}}\right)(m_Y(X)-\theta).
\end{align*}
Notice that $\psi^{*}_{CS-2}(W;\theta)\in\mathcal{T}$. Also for any $D=d$, $T=t$ we obtain
\begin{align*}
\mathbb{E}&\left[\frac{p_{D}(X)p_{T}(X)}{p_{DT}}\frac{G_{d,t}}{p_{D=d}(X)p_{T=t}(X)}(Y-m_Y(d,t,X))\times S(Y,D,T,X;\beta_0)\right]\\
&=\frac{1}{p_{DT}}\int\int yS_y(d,t,x;\beta_0)f_{Y|D,T,X}(y|d,t,x)p_{D}(x)p_T(x)f_X(x)dydx.
\end{align*}
Further,
\begin{align*}
&\mathbb{E}\left[\frac{p_T(X)D}{p_{DT}}(m_Y(X)-\theta)\times S(Y,D,T,X;\beta_0)\right]
=\frac{1}{p_{DT}}\mathbb{E}\left[(m_Y(X)-\theta)\left(\dot{p}_{D}(X;\beta_0)p_T(X)+p_{D}(X)p_T(X)S_x(X;\beta_0)\right)\right]\\
&\mathbb{E}\left[\frac{p_D(X)T}{p_{DT}}(m_Y(X)-\theta)\times S(Y,D,T,X;\beta_0)\right]
=\frac{1}{p_{DT}}\mathbb{E}\left[(m_Y(X)-\theta)\left(p_D(X)\dot{p}_{T}(X;\beta_0)+p_{D}(X)p_T(X)S_x(X;\beta_0)\right)\right]\\
&\mathbb{E}\left[\frac{p_D(X)p_T(X)}{p_{DT}}(m_Y(X)-\theta)\times S(Y,D,T,X;\beta_0)\right]
=\frac{1}{p_{DT}}\mathbb{E}\left[(m_Y(X)-\theta)\left(p_{D}(X)p_T(X)S_x(X;\beta_0)\right)\right]
\end{align*}
such that
\begin{align*}
\mathbb{E}&\left[\left(\frac{p_T(X)(D-p_D(X))}{p_{DT}}+\frac{p_D(X)(T-p_T(X))}{p_{DT}}+\frac{p_D(X)p_T(X)}{p_{DT}}\right)(m_Y(X)-\theta)\times S(Y,D,T,X;\beta_0)\right]\\
&=\frac{1}{p_{DT}}\int(m_Y(X)-\theta)\left(\dot{p}_{D}(x;\beta_0)p_T(x)+p_{D}(x)\dot{p}_T(x;\beta_0)+p_D(x)p_T(x)S_x(x;\beta_0)\right)f_X(x)dx.
\end{align*}
It follows that $\psi^{*}_{CS-2}(W;\theta)$ is the efficient influence function and the variance lower bound for (CS-2) is $\mathbb{E}[\psi^{*}_{CS-2}(W;\theta)^2]$.
\subsubsection*{Proof for CS-3}
Under the conditions in (CS-3)
\begin{align*}
S_{d,t}(d,t,x;\beta_0)&=t\left(\frac{d}{p_D(1,x)}-\frac{1-d}{1-p_D(1,x)}\right)\dot{p}_D(1,x;\beta_0)+(1-t)\left(\frac{d}{p_D(0,x)}-\frac{1-d}{1-p_D(0,x)}\right)\dot{p}_D(0,x;\beta_0)\\
&+\left(\frac{t}{p_T}-\frac{1-t}{1-p_T}\right)\dot{p}_T(\beta_0).
\end{align*}
The tangent space of the model is characterized by the set of functions that are mean zero and satisfy the structure of the score function
\begin{align*}
\mathcal{T}&=\Bigg\{\sum_{d=0}^1\sum_{t=0}^1(g_{d,t}S_y(d,t,x))+t(d-p_D(1,x))a(x)+(1-t)(d-p_D(0,x))b(x)+(t-p_T)c+S_x(x)\Bigg\}
\end{align*}
for any functions $S_y(d,t,x)$ and $S_x(x)$ that satisfy
\begin{align*}
\mathbb{E}&\left[S_y(d,t,X)|D=d,T=t,X\right]=0\\
\mathbb{E}&\left[S_x(X)\right]=0
\end{align*}
and any square integrable functions $a(x)$ and $b(x)$ and some constant $c$.\\
The parameter $\theta$ is pathwise differentiable. For the parametric submodel we have
\begin{align*}
\theta(\beta)=\frac{\sum_{d=0}^1\sum_{t=0}^1(-1)^{(d+t)}\int\int y f_{Y|D,T,X}(y|d,t,x;\beta)p_{D}(1,x;\beta)f_X(x;\beta)dydx}{\int p_{D}(1,x;\beta)f_X(x;\beta)dx}
\end{align*}
and at $\beta=\beta_0$
\begin{align*}
\frac{\partial \theta(\beta_0)}{\partial \beta}&=\frac{1}{p_{D}(1)}\left(\sum_{d=0}^1\sum_{t=0}^1(-1)^{(d+t)}\int\int yS_y(d,t,x;\beta_0)f_{Y|D,T,X}(y|d,t,x)p_{D}(1,x)f_X(x)dydx\right)\\
&+\frac{1}{p_{D}(1)}\int(m_Y(X)-\theta)\left(\dot{p}_{D}(1,x;\beta_0)+p_D(1,x)S_x(x;\beta_0)\right)f_X(x)dx.
\end{align*}
We consider the function
\begin{align*}
\psi^{*}_{CS-3}(W;\theta)&=\frac{p_{D}(1,X)}{p_{D}(1)}\left(\sum_{d=0}^1\sum_{t=0}^1(-1)^{(d+t)}\frac{G_{d,t}}{p_{D=d}(t,X)p_{T=t}}(Y-m_Y(d,t,X))\right)\\
&+\left(\frac{T(D-p_D(1,X))}{p_{D}(1)p_T}+\frac{p_D(1,X)}{p_{D}(1)}\right)(m_Y(X)-\theta).
\end{align*}
Notice that $\psi^{*}_{CS-3}(W;\theta)\in\mathcal{T}$. Also for any $D=d$, $T=t$ we obtain
\begin{align*}
\mathbb{E}&\left[\frac{p_{D}(1,X)}{p_{D}(1)}\frac{G_{d,t}}{p_{D=d}(t,X)p_{T=t}}(Y-m_Y(d,t,X))\times S(Y,D,T,X;\beta_0)\right]\\
&=\frac{1}{p_{D}(1)}\int\int yS_y(d,t,x;\beta_0)f_{Y|D,T,X}(y|d,t,x)p_{D}(1,x)f_X(x)dydx.
\end{align*}
Further,
\begin{align*}
&\mathbb{E}\left[\frac{DT}{p_{D}(1)p_T}(m_Y(X)-\theta)\times S(Y,D,T,X;\beta_0)\right]\\
&=\frac{1}{p_{D}(1)}\mathbb{E}\left[(m_Y(X)-\theta)\left(\dot{p}_{D}(1,X;\beta_0)+p_D(1,X)\frac{\dot{p}_T(\beta_0)}{p_T}+p_{D}(1,X)S_x(X;\beta_0)\right)\right]\\
&\mathbb{E}\left[\frac{p_D(1,X)T}{p_{D}(1)}(m_Y(X)-\theta)\times S(Y,D,T,X;\beta_0)\right]
=\frac{1}{p_{D}(1)}\mathbb{E}\left[(m_Y(X)-\theta)\left(p_D(1,X)\frac{\dot{p}_T(\beta_0)}{p_T}+p_{D}(1,X)S_x(X;\beta_0)\right)\right]\\
&\mathbb{E}\left[\frac{p_D(1,X)}{p_{D}(1)}(m_Y(X)-\theta)\times S(Y,D,T,X;\beta_0)\right]
=\frac{1}{p_{D}(1)}\mathbb{E}\left[(m_Y(X)-\theta)\left(p_{D}(1,X)S_x(X;\beta_0)\right)\right]
\end{align*}
such that
\begin{align*}
\mathbb{E}&\left[\left(\frac{T(D-p_D(1,X))}{p_{D}(1)p_T}+\frac{p_D(1,X)}{p_{D}(1)}\right)(m_Y(X)-\theta)\times S(Y,D,T,X;\beta_0)\right]\\
&=\frac{1}{p_{D}(1)}\int(m_Y(X)-\theta)\left(\dot{p}_{D}(1,x;\beta_0)+p_D(1,x)S_x(x;\beta_0)\right)f_X(x)dx.
\end{align*}
It follows that $\psi^{*}_{CS-3}(W;\theta)$ is the efficient influence function and the variance lower bound for (CS-3) is $\mathbb{E}[\psi^{*}_{CS-3}(W;\theta)^2]$.
\subsubsection*{Proof for CS-4}
A similar proof can be found in \textcite{SantAnna_Zhao_2020}. We give it here for completeness.\\
Under the conditions in (CS-4)
\begin{align*}
S_{d,t}(d,t,x;\beta_0)=\left(\frac{d}{p_D(x)}-\frac{1-d}{1-p_D(x)}\right)\dot{p}_D(x;\beta_0)+\left(\frac{t}{p_T}-\frac{1-t}{1-p_T}\right)\dot{p}_T(\beta_0).
\end{align*}
The tangent space of the model is characterized by the set of functions that are mean zero and satisfy the structure of the score function
\begin{align*}
\mathcal{T}&=\Bigg\{\sum_{d=0}^1\sum_{t=0}^1(g_{d,t}S_y(d,t,x))+(d-p_D(x))a(x)+(t-p_T)b+S_x(x)\Bigg\}
\end{align*}
for any functions $S_y(d,t,x)$ and $S_x(x)$ that satisfy
\begin{align*}
\mathbb{E}&\left[S_y(d,t,X)|D=d,T=t,X\right]=0\\
\mathbb{E}&\left[S_x(X)\right]=0
\end{align*}
and any square integrable functions $a(x)$ and some constant $b$.\\
The parameter $\theta$ is pathwise differentiable. For the parametric submodel we have
\begin{align*}
\theta(\beta)=\frac{\sum_{d=0}^1\sum_{t=0}^1(-1)^{(d+t)}\int\int y f_{Y|D,T,X}(y|d,t,x;\beta)p_{D}(x;\beta)f_X(x;\beta)dydx}{\int p_{D}(x;\beta)f_X(x;\beta)dx}
\end{align*}
and at $\beta=\beta_0$
\begin{align*}
\frac{\partial \theta(\beta_0)}{\partial \beta}&=\frac{1}{p_{D}}\left(\sum_{d=0}^1\sum_{t=0}^1(-1)^{(d+t)}\int\int yS_y(d,t,x;\beta_0)f_{Y|D,T,X}(y|d,t,x)p_{D}(x)f_X(x)dydx\right)\\
&+\frac{1}{p_{D}}\int(m_Y(X)-\theta)\left(\dot{p}_{D}(x;\beta_0)+p_D(x)S_x(x;\beta_0)\right)f_X(x)dx.
\end{align*}
We consider the function
\begin{align*}
\psi^{*}_{CS-4}(W;\theta)=\frac{p_{D}(X)}{p_{D}}\left(\sum_{d=0}^1\sum_{t=0}^1(-1)^{(d+t)}\frac{G_{d,t}}{p_{D=d}(X)p_{T=t}}(Y-m_Y(d,t,X))\right)+\frac{D}{p_D}(m_Y(X)-\theta).
\end{align*}
Notice that $\psi^{*}_{CS-4}(W;\theta)\in\mathcal{T}$. Also for any $D=d$, $T=t$ we obtain
\begin{align*}
\mathbb{E}&\left[\frac{p_{D}(X)}{p_{D}}\frac{G_{d,t}}{p_{D=d}(X)p_{T=t}}(Y-m_Y(d,t,X))\times S(Y,D,T,X;\beta_0)\right]\\
&=\frac{1}{p_{D}}\int\int yS_y(d,t,x;\beta_0)f_{Y|D,T,X}(y|d,t,x)p_{D}(x)f_X(x)dydx.
\end{align*}
Further,
\begin{align*}
\mathbb{E}\left[\frac{D}{p_{D}}(m_Y(X)-\theta)\times S(Y,D,T,X;\beta_0)\right]=\frac{1}{p_{D}}\mathbb{E}\left[(m_Y(X)-\theta)\left(\dot{p}_{D}(X;\beta_0)+p_D(X)S_x(X;\beta_0)\right)\right].
\end{align*}
It follows that $\psi^{*}_{CS-4}(W;\theta)$ is the efficient influence function and the variance lower bound for (CS-4) is $\mathbb{E}[\psi^{*}_{CS-4}(W;\theta)^2]$.
\subsubsection*{Proof for CS-5}
Under the conditions in (CS-5)
\begin{align*}
S_{d,t}(d,t,x;\beta_0)=\left(\frac{d}{p_D}-\frac{1-d}{1-p_D}\right)\dot{p}_D(\beta_0)+\left(\frac{t}{p_T}-\frac{1-t}{1-p_T}\right)\dot{p}_T(\beta_0).
\end{align*}
The tangent space of the model is characterized by the set of functions that are mean zero and satisfy the structure of the score function
\begin{align*}
\mathcal{T}&=\Bigg\{\sum_{d=0}^1\sum_{t=0}^1(g_{d,t}S_y(d,t,x))+(d-p_D)a+(t-p_T)b+S_x(x)\Bigg\}
\end{align*}
for any functions $S_y(d,t,x)$ and $S_x(x)$ that satisfy
\begin{align*}
\mathbb{E}&\left[S_y(d,t,X)|D=d,T=t,X\right]=0\\
\mathbb{E}&\left[S_x(X)\right]=0
\end{align*}
and the constants $a$ and $b$.\\
The parameter $\theta$ is pathwise differentiable. For the parametric submodel we have
\begin{align*}
\theta(\beta)=\sum_{d=0}^1\sum_{t=0}^1(-1)^{(d+t)}\int\int y f_{Y|D,T,X}(y|d,t,x;\beta)f_X(x;\beta)dydx
\end{align*}
and at $\beta=\beta_0$
\begin{align*}
\frac{\partial \theta(\beta_0)}{\partial \beta}&=\sum_{d=0}^1\sum_{t=0}^1(-1)^{(d+t)}\int\int yS_y(d,t,x;\beta_0)f_{Y|D,T,X}(y|d,t,x)f_X(x)dydx\\
&+\sum_{d=0}^1\sum_{t=0}^1(-1)^{(d+t)}\int\int yf_{Y|D,T,X}(y|d,t,x)S_x(X;\beta_0)f_X(x)dydx.
\end{align*}
We consider the function
\begin{align*}
\psi^{*}_{CS-5}(W;\theta)=\sum_{d=0}^1\sum_{t=0}^1(-1)^{(d+t)}\frac{G_{d,t}}{p_{D=d}p_{T=t}}(Y-m_Y(d,t,X))+m_Y(X)-\theta.
\end{align*}
Notice that $\psi^{*}_{CS-5}(W;\theta)\in\mathcal{T}$. Also for any $D=d$, $T=t$ we obtain
\begin{align*}
\mathbb{E}&\left[\frac{G_{d,t}}{p_{D=d}p_{T=t}}(Y-m_Y(d,t,X))\times S(Y,D,T,X;\beta_0)\right]\\
&=\int\int yS_y(d,t,x;\beta_0)f_{Y|D,T,X}(y|d,t,x)f_X(x)dydx.
\end{align*}
Further,
\begin{align*}
\mathbb{E}\left[(m_Y(X)-\theta)\times S(Y,D,T,X;\beta_0)\right]=\mathbb{E}\left[m_Y(X)S_x(X;\beta_0)\right].
\end{align*}
It follows that $\psi^{*}_{CS-5}(W;\theta)$ is the efficient influence function and the variance lower bound for (CS-5) is $\mathbb{E}[\psi^{*}_{CS-5}(W;\theta)^2]$.
\subsection{Proof of Corollary \ref{cor:releffcs1}}\label{app:releffcs1}
From Theorem \ref{thm:effcs} we directly obtain
\begin{align*}
&\mathbb{E}\left[\psi^{*}_{CS}(W;\theta)^2\right]=\mathbb{E}\Bigg[\frac{q_{CS;D=1,T=1}(X)^2}{q_{CS;DT}^2}\Bigg(\mathbb{E}\left[\psi^{a}_{CS}(W)^2|X\right]+\frac{q_{CS;DT}^2}{q_{CS;D=1,T=1}(X)^2}\mathbb{E}\left[\psi^{b}_{CS}(W)^2|X\right]\left(m_Y(X)-\theta\right)^2\\
&+2\frac{q_{CS;DT}}{q_{CS;D=1,T=1}(X)}\mathbb{E}\left[\psi^{a}_{CS}(W)\psi^{b}_{CS}(W)|X\right]\left(m_Y(X)-\theta\right)\Bigg)\Bigg]
\end{align*}
where
\begin{align*}
&\mathbb{E}\left[\psi^{a}_{CS}(W)^2|X\right]=\mathbb{E}\left[\sum_{d=0}^1\sum_{t=0}^1\frac{\text{Var}(Y|D=d,T=t,X)}{q_{CS;D=d,T=t(X)}}\right]\\
&\mathbb{E}\left[\psi^{a}_{CS}(W)\psi^{b}_{CS}(W)|X\right]=0
\end{align*}
for all settings (CS-1) to (CS-5). Further, we obtain
\begin{align*}
&\mathbb{E}\left[\psi^{b}_{CS-1}(W)^2|X\right]=\frac{p_{D=1,T=1}(X)^2}{p_{DT}^2}\frac{1}{p_{D=1,T=1}(X)}\\
&\mathbb{E}\left[\psi^{b}_{CS-2}(W)^2|X\right]=\frac{p_{D}(X)^2p_T(X)^2}{p_{DT}^2}\left(\frac{1}{p_D(X)}+\frac{1}{p_T(X)}-1\right)\\
&\mathbb{E}\left[\psi^{b}_{CS-3}(W)^2|X\right]=\frac{p_{D}(1,X)^2}{p_{D}(1)^2}p_T\left(\frac{1}{p_D(1,X)}+\frac{1}{p_T}-1\right)\\
&\mathbb{E}\left[\psi^{b}_{CS-4}(W)^2|X\right]=\frac{p_{D}(X)^2}{p_{D}^2}\frac{1}{p_D(X)}\\
&\mathbb{E}\left[\psi^{b}_{CS-5}(W)^2|X\right]=1.
\end{align*}
Suppose that any of the assumptions in settings (CS-2)-(CS-5) is true then the efficiency bound of (CS-5) is higher by
\begin{align*}
\Delta_{CS-1,CS}=\mathbb{E}\left[\frac{q_{CS;D=1,T=1}(X)^2}{q_{CS;DT}^2}\left(m_Y(X)-\theta\right)^2\left(\frac{1}{q_{CS;D=1,T=1}(X)}-\frac{q_{CS;DT}^2}{q_{CS;D=1,T=1}(X)^2}\mathbb{E}\left[\psi^{b}_{CS}(W)^2|X\right]\right)\right].
\end{align*}
Similar arguments can be made for the comparison of the other bounds.
\subsection{Proof of Theorem \ref{thm:effpa}}\label{app:effpa}
We observe the data $W=(Y(0),Y(1),D,X)$. For the distribution of the data consider a regular parametric submodel indexed by $\beta$. The density under the submodel can then be written as
\begin{align*}
f_W(w;\beta)=f_{Y(0),Y(1)|D,X}(y(0),y(1)|d,x;\beta)f_{D|X}(d|x;\beta)f_{X}(x;\beta)
\end{align*}
which equals $f_W(w)$ at $\beta=\beta_0$.\\
The score function is defined as $S(y(0),y(1),d,x;\beta_0)=\frac{\partial \log f_W(w;\beta_0)}{\partial \beta}$ and we obtain $S(y(0),y(1),d,x;\beta_0)=S_{y(0),y(1)}(y(0),y(1),d,x;\beta_0)+S_{d}(d,x;\beta_0)+S_x(x;\beta_0)$ with
\begin{align*}
S&_{y(0),y(1)}(d,x;\beta_0)=\frac{\partial \log f_{Y(0),Y(1)|D,X}(y(0),y(1)|d,x;\beta_0)}{\partial \beta}\\
S&_{d}(d,x;\beta_0)=\frac{\partial \log f_{D|X}(d|x;\beta_0)}{\partial \beta}\\
S&_x(x;\beta_0)=\frac{\partial \log f_{X}(x;\beta_0)}{\partial \beta}
\end{align*}
where 
\begin{align*}
S_{y(0),y(1)}(y(0),y(1),d,x;\beta_0)&=dS_{y(0),y(1)}(1,x;\beta_0)+(1-d)S_{y(0),y(1)}(0,x;\beta_0)
\end{align*}
and $S_{d}(d,x;\beta_0)$ again depends on the assumptions (PA-1) and (PA-2).\\~\\
For all regular parametric submodels the variance lower bound for a model is the second moment of the projection of a function $\psi^{*}_{PA}(W;\theta)$ (with $\mathbb{E}[\psi^{*}_{PA}(W;\theta)]=0$ and an existing second moment) on the tangent space $\mathcal{T}$ that satisfies
\begin{align*}
\frac{\partial \theta(\beta_0)}{\partial \beta}=\mathbb{E}\left[\psi^{*}_{PA}(W;\theta)S(Y(0),Y(1),D,X;\beta_0)\right].
\end{align*}
When $\psi^{*}_{PA}(W;\theta)\in\mathcal{T}$, the projection on $\mathcal{T}$ is the function itself and therefore the variance lower bound for the model is given by $\mathbb{E}[\psi^{*}_{PA}(W;\theta)^2]$.
\subsubsection*{Proof for PA-1}
Under the conditions in (PA-1)
\begin{align*}
S_{d}(d,x;\beta_0)=\left(\frac{d}{p_D(x)}-\frac{1-d}{1-p_D(x)}\right)\dot{p}_D(x;\beta_0).
\end{align*}
The tangent space of the model is characterized by the set of functions that are mean zero and satisfy the structure of the score function
\begin{align*}
\mathcal{T}&=\Bigg\{dS_{y(0),y(1)}(1,x))+(1-d)S_{y(0),y(1)}(0,x))+(d-p_D(x))a(x)+S_x(x)\Bigg\}
\end{align*}
for any functions $S_{y(0),y(1)}(d,x)$ and $S_x(x)$ that satisfy
\begin{align*}
\mathbb{E}&\left[S_{y(0),y(1)}(d,X)|D=d,X\right]=0\\
\mathbb{E}&\left[S_x(X)\right]=0
\end{align*}
and any square integrable function $a(x)$.\\
The parameter $\theta$ is pathwise differentiable. For the parametric submodel we have
\begin{align*}
\theta(\beta)&=\frac{\int\int\int \left(y(1)-y(0)\right) f_{Y(0),Y(1)|D,X}(y(0),y(1)|1,x;\beta)p_{D}(x;\beta)f_X(x;\beta)dy(0)dy(1)dx}{\int p_{D}(x;\beta)f_X(x;\beta)dx}\\
&-\frac{\int\int\int \left(y(1)-y(0)\right) f_{Y(0),Y(1)|D,X}(y(0),y(1)|0,x;\beta)p_{D}(x;\beta)f_X(x;\beta)dy(0)dy(1)dx}{\int p_{D}(x;\beta)f_X(x;\beta)dx}
\end{align*}
and at $\beta=\beta_0$
\begin{align*}
\frac{\partial \theta(\beta_0)}{\partial \beta}&=\frac{1}{p_{D}}\int\int\int \left(y(1)-y(0)\right)S_{y(0),y(1)}(1,x;\beta_0)f_{Y(0),Y(1)|D,X}(y(0),y(1)|1,x)p_{D}(x)f_X(x)dy(0)dy(1)dx\\
&-\frac{1}{p_{D}}\int\int\int \left(y(1)-y(0)\right)S_{y(0),y(1)}(0,x;\beta_0)f_{Y(0),Y(1)|D,X}(y(0),y(1)|0,x)p_{D}(x)f_X(x)dy(0)dy(1)dx\\
&+\frac{1}{p_{D}}\int(m_{\Delta Y}(x)-\theta)\left(\dot{p}_{D}(x;\beta_0)+p_{D}(x)S_x(x;\beta_0)\right)f_X(x)dx.
\end{align*}
We consider the function
\begin{align*}
\psi^{*}_{PA-1}(W;\theta)&=\frac{p_{D}(X)}{p_{D}}\left(\frac{D}{p_{D}(X)}(Y(1)-Y(0)-m_{\Delta Y}(1,X))-\frac{1-D}{1-p_{D}(X)}(Y(1)-Y(0)-m_{\Delta Y}(0,X))\right)\\
&+\frac{D}{p_D}\left(m_{\Delta Y}(X)-\theta\right).
\end{align*}
Notice that $\psi^{*}_{PA-1}(W;\theta)\in\mathcal{T}$. Also we obtain
\begin{align*}
\mathbb{E}&\left[\frac{p_{D}(X)}{p_{D}}\frac{D}{p_{D}(X)}(Y(1)-Y(0)-m_{\Delta Y}(1,X))\times S(Y,D,X;\beta_0)\right]\\
&=\frac{1}{p_{D}}\int\int\int \left(y(1)-y(0)\right)S_{y(0),y(1)}(1,x;\beta_0)f_{Y(0),Y(1)|D,X}(y(0),y(1)|1,x)p_{D}(x)f_X(x)dy(0)dy(1)dx\\
\mathbb{E}&\left[\frac{p_{D}(X)}{p_{D}}\frac{1-D}{1-p_{D}(X)}(Y(1)-Y(0)-m_{\Delta Y}(0,X))\times S(Y,D,X;\beta_0)\right]\\
&=\frac{1}{p_{D}}\int\int\int \left(y(1)-y(0)\right)S_{y(0),y(1)}(0,x;\beta_0)f_{Y(0),Y(1)|D,X}(y(0),y(1)|0,x)p_{D}(x)f_X(x)dy(0)dy(1)dx.
\end{align*}
Further,
\begin{align*}
&\mathbb{E}\left[\frac{D}{p_{D}}(m_{\Delta Y}(X)-\theta)\times S(Y,D,X;\beta_0)\right]
=\frac{1}{p_{D}}\int(m_{\Delta Y}(x)-\theta)\left(\dot{p}_{D}(x;\beta_0)+p_{D}(x)S_x(x;\beta_0)\right)f_X(x)dx.
\end{align*}
It follows that $\psi^{*}_{PA-1}(W;\theta)$ is the efficient influence function and the variance lower bound for (PA-1) is $\mathbb{E}[\psi^{*}_{PA-1}(W;\theta)^2]$.
\subsubsection*{Proof for PA-2}
Under the conditions in (PA-2)
\begin{align*}
S_{d}(d,x;\beta_0)=\left(\frac{d}{p_D}-\frac{1-d}{1-p_D}\right)\dot{p}_D(\beta_0).
\end{align*}
The tangent space of the model is characterized by the set of functions that are mean zero and satisfy the structure of the score function
\begin{align*}
\mathcal{T}&=\Bigg\{dS_{y(0),y(1)}(1,x))+(1-d)S_{y(0),y(1)}(0,x))+(d-p_D)a+S_x(x)\Bigg\}
\end{align*}
for any functions $S_{y(0),y(1)}(d,x)$ and $S_x(x)$ that satisfy
\begin{align*}
\mathbb{E}&\left[S_{y(0),y(1)}(d,X)|D=d,X\right]=0\\
\mathbb{E}&\left[S_x(X)\right]=0
\end{align*}
and a constant $a$.\\
The parameter $\theta$ is pathwise differentiable. For the parametric submodel we have
\begin{align*}
\theta(\beta)&=\int\int\int \left(y(1)-y(0)\right) f_{Y(0),Y(1)|D,X}(y(0),y(1)|1,x;\beta)f_X(x;\beta)dy(0)dy(1)dx\\
&-\int\int\int \left(y(1)-y(0)\right) f_{Y(0),Y(1)|D,X}(y(0),y(1)|0,x;\beta)f_X(x;\beta)dy(0)dy(1)dx
\end{align*}
and at $\beta=\beta_0$
\begin{align*}
\frac{\partial \theta(\beta_0)}{\partial \beta}&=\int\int\int \left(y(1)-y(0)\right)S_{y(0),y(1)}(1,x;\beta_0)f_{Y(0),Y(1)|D,X}(y(0),y(1)|1,x)f_X(x)dy(0)dy(1)dx\\
&-\int\int\int \left(y(1)-y(0)\right)S_{y(0),y(1)}(0,x;\beta_0)f_{Y(0),Y(1)|D,X}(y(0),y(1)|0,x)f_X(x)dy(0)dy(1)dx\\
&+\int m_{\Delta Y}(x)S_x(x;\beta_0)f_X(x)dx.
\end{align*}
We consider the function
\begin{align*}
\psi^{*}_{PA-2}(W;\theta)&=\frac{D}{p_{D}}(Y(1)-Y(0)-m_{\Delta Y}(1,X))-\frac{1-D}{1-p_{D}}(Y(1)-Y(0)-m_{\Delta Y}(0,X))+m_{\Delta Y}(X)-\theta.
\end{align*}
Notice that $\psi^{*}_{PA-2}(W;\theta)\in\mathcal{T}$. Similarly to the proof for (PA-1) it can be shown that
\begin{align*}
\frac{\partial \theta(\beta_0)}{\partial \beta}=\mathbb{E}\left[\psi^{*}_{PA-2}(W;\theta)\times S(Y,D,X;\beta_0)\right].
\end{align*}
It follows that $\psi^{*}_{PA-2}(W;\theta)$ is the efficient influence function and the variance lower bound for (PA-2) is $\mathbb{E}[\psi^{*}_{PA-2}(W;\theta)^2]$.
\subsection{Proof of Corollary \ref{cor:releffcspa}}\label{app:releffcspa}
Notice that for cross-sectional data we can write $Y=TY(1)+(1-T)Y(0)$ and we have $m_{Y}(d,t,X)=\mathbb{E}\left[Y(t)|D=d,X\right]$. Further, under the conditions of a panel for case (CS-1) we obtain $p_{D=d,T=t}(X)=\frac{1}{2}p_{D}(X)$. Hence, when we observe the panel structure the efficiency bound for (CS-1) is
\begin{align*}
\mathbb{E}\left[2\frac{p_D(X)^2}{p_D^2}\left(\sum_{d=0}^1\sum_{t=0}^1\frac{\text{Var}(Y(t)|D=d,X)}{p_{D=d}(X)}+\frac{(m_{\Delta Y}(X)-\theta)^2}{p_D(X)}\right)\right].
\end{align*}
For 
\begin{align*}
\Delta_{CS-1,PA-1}=\mathbb{E}\left[2\frac{p_D(X)^2}{p_D^2}\left(\sum_{d=0}^1\sum_{t=0}^1\frac{\text{Var}(Y(t)|D=d,X)}{p_{D=d}(X)}+\frac{(m_{\Delta Y}(X)-\theta)^2}{p_D(X)}\right)\right]-\mathbb{E}\left[\psi^{*}_{PA-1}(W;\theta)^2\right]
\end{align*}
the first claim immediately follows.\\
The second claim follows similarly by observing that when we observe the panel structure the efficiency bound for (CS-5) is
\begin{align*}
\mathbb{E}\left[2\sum_{d=0}^1\sum_{t=0}^1\frac{\text{Var}(Y(t)|D=d,X)}{p_{D=d}}+(m_{\Delta Y}(X)-\theta)^2\right].
\end{align*}
\section{Proofs for Section \ref{sec:estinf}}
\subsection{Lemma \ref{lm:firststage}}
The general result of the lemma can be similarly found for example in \textcite{Chernozhukov_Chetverikov_Demirer_Duflo_Hansen_Newey_2017}.
\begin{lemma}\label{lm:firststage}
For a generic function $\psi(W,\cdot)$ let $\psi(W,\eta)$ be the version with true nuisance parameters and $\psi(W,\hat{\eta}_{-k})$ the version with cross-fitted nuisance parameters. Define $\bar{\psi}=\psi(W,\hat{\eta}_{-k})-\psi(W,\eta)$. Then under the conditions that (i) $\mathbb{E}\left[\bar{\psi}\right]=o_p\left(N^{-\frac{1}{2}}\right)$ and (ii) $\left\lVert\bar{\psi}\right\rVert_2=o_p(1)$ the term $\frac{1}{N}\sum_{k=1}^K\sum_{i\in\mathcal{I}^k}^n\bar{\psi}_i$ is bounded such that
\begin{align*}
\left\lVert\frac{1}{N}\sum_{k=1}^K\sum_{i\in\mathcal{I}^k}^n\bar{\psi}_i\right\rVert_2=o_p\left(N^{-\frac{1}{2}}\right).
\end{align*} 
\end{lemma}
\textit{Proof:}\\
Write
\begin{align*}
\left\lVert\frac{1}{N}\sum_{k=1}^K\sum_{i\in\mathcal{I}^k}^n\bar{\psi}_i\right\rVert_2\leq\left\lVert\mathbb{E}\left[\bar{\psi}\right]\right\rVert_2+\left\lVert\frac{1}{N}\sum_{k=1}^K\sum_{i\in\mathcal{I}^k}^n\left(\bar{\psi}_i-\mathbb{E}\left[\bar{\psi}\right]\right)\right\rVert_2
\end{align*}
and notice that for every $k\in [1,...,K]$
\begin{align*}
\mathbb{E}\left[\left\lVert\frac{1}{\sqrt{n}}\sum_{i\in\mathcal{I}^k}^n\left(\bar{\psi}_i-\mathbb{E}\left[\bar{\psi}\right]\right)\right\rVert_2^2\right]&=\mathbb{E}\left[\left\lVert\frac{1}{\sqrt{n}}\sum_{i\in\mathcal{I}^k}^n\left(\bar{\psi}_i-\mathbb{E}\left[\bar{\psi}\right]\right)\right\rVert_2^2\Bigg|W_{i\in\mathcal{I}^k}\right]\\
&\leq\mathbb{E}\left[\left(\frac{1}{\sqrt{n}}\sum_{i\in\mathcal{I}^k}^n\left(\bar{\psi}_i-\mathbb{E}\left[\bar{\psi}\right]\right)\right)^2\Bigg|W_{i\in\mathcal{I}^k}\right]\\
&\leq\sup_{\hat{\eta}}\mathbb{E}\left[\left(\psi(W,\hat{\eta}_{-k})-\psi(W,\eta)\right)^2\right]
\end{align*}
such that
\begin{align*}
\left\lVert\frac{1}{N}\sum_{k=1}^K\sum_{i\in\mathcal{I}^k}^n\left(\bar{\psi}_i-\mathbb{E}\left[\bar{\psi}\right]\right)\right\rVert_2\leq\frac{1}{\sqrt{N}}\left\lVert\bar{\psi}\right\rVert_2.
\end{align*}
Thus, $\sqrt{N}\mathbb{E}\left[\bar{\psi}\right]=o_p(1)$ and $\sqrt{N}\left\lVert\frac{1}{N}\sum_{k=1}^K\sum_{i\in\mathcal{I}^k}^n\left(\bar{\psi}_i-\mathbb{E}\left[\bar{\psi}\right]\right)\right\rVert_2=o_p(1)$ under the conditions (i) and (ii).
\subsection{Proof of Theorem \ref{thm:estinf}}\label{app:estinf}
We have to show that
\begin{align*}
\left\lVert\sqrt{N}\left(\hat{\theta}-\theta\right)-\frac{1}{\sqrt{N}}\sum_{i=1}^N\psi(W_i,\eta;\theta)\right\rVert_2=o_p(1).
\end{align*}
This implies that $\hat{\theta}$ is asymptotically linear with influence function $\psi(W,\eta;\theta)$. Thus, given the above holds, the Lindeberg-L\'{e}vy Central Limit Theorem implies the second claim of the Theorem.\\
In order to derive the result write
\begin{align*}
\hat{\theta}-\theta&=\left(\frac{1}{N}\sum_{k=1}^K\sum_{i\in\mathcal{I}^k}^n\psi^b(W_i,\hat{\eta}_{-k})\right)^{-1}\times\frac{1}{N}\sum_{k=1}^K\sum_{i\in\mathcal{I}^k}^n\psi(W_i,\hat{\eta}_{-k};\theta)\\
&=\frac{1}{N}\sum_{k=1}^K\sum_{i\in\mathcal{I}^k}^n\psi(W_i,\hat{\eta}_{-k};\theta)+\left(\frac{1}{\frac{1}{N}\sum_{k=1}^K\sum_{i\in\mathcal{I}^k}^n\psi^b(W_i,\eta)}-1\right)\times\frac{1}{N}\sum_{k=1}^K\sum_{i\in\mathcal{I}^k}^n\psi(W_i,\hat{\eta}_{-k};\theta)\\
&+\left(\frac{1}{\frac{1}{N}\sum_{k=1}^K\sum_{i\in\mathcal{I}^k}^n\psi^b(W_i,\hat{\eta}_{-k})}-\frac{1}{\frac{1}{N}\sum_{k=1}^K\sum_{i\in\mathcal{I}^k}^n\psi^b(W_i,\eta)}\right)\times\frac{1}{N}\sum_{k=1}^K\sum_{i\in\mathcal{I}^k}^n\psi(W_i,\hat{\eta}_{-k};\theta)
\end{align*}
and define
\begin{align*}
&i=\frac{1}{N}\sum_{k=1}^K\sum_{i\in\mathcal{I}^k}^n\left(\psi(W_i,\hat{\eta}_{-k};\theta)-\psi(W_i,\eta;\theta)\right)\\
&ii=\left(\frac{1}{\frac{1}{N}\sum_{k=1}^K\sum_{i\in\mathcal{I}^k}^n\psi^b(W_i,\eta)}-1\right)\\
&iii=\left(\frac{1}{\frac{1}{N}\sum_{k=1}^K\sum_{i\in\mathcal{I}^k}^n\psi^b(W_i,\hat{\eta}_{-k})}-\frac{1}{\frac{1}{N}\sum_{k=1}^K\sum_{i\in\mathcal{I}^k}^n\psi^b(W_i,\eta)}\right).
\end{align*}
Suppose that $\left\lVert i\right\rVert_2=o_p\left(N^{-\frac{1}{2}}\right)$, $\left\lvert ii\right\rvert=O_p\left(N^{-\frac{1}{2}}\right)$, $\left\lvert iii\right\rvert=o_p\left(N^{-\frac{1}{2}}\right)$ and notice that $\left\lVert\frac{1}{\sqrt{N}}\sum_{i=1}^N\psi(W_i,\eta;\theta)\right\rVert_2=O_p(1)$ and
\begin{align*}
\sqrt{N}\left(\hat{\theta}-\theta\right)-\frac{1}{\sqrt{N}}\sum_{i=1}^N\psi(W_i,\eta;\theta)=\sqrt{N}i+\left(\frac{1}{\sqrt{N}}\sum_{i=1}^N\psi(W_i,\eta;\theta)+\sqrt{N}i\right)\times\left(ii+iii\right).
\end{align*}
We obtain
\begin{align*}
\left\lVert\sqrt{N}\left(\hat{\theta}-\theta\right)-\frac{1}{\sqrt{N}}\sum_{i=1}^N\psi(W_i,\eta;\theta)\right\rVert_2&\leq o_p(1)+O_p(1)O_p\left(N^{-\frac{1}{2}}\right)+O_p(1)o_p\left(N^{-\frac{1}{2}}\right)+o_p(1)O_p\left(N^{-\frac{1}{2}}\right)\\
&+o_p(1)o_p\left(N^{-\frac{1}{2}}\right)\\
&=o_p(1).
\end{align*}
In the following we will derive the bounds for the different terms.\\~\\
\textit{Bounding i}\\
Notice that
\begin{align*}
&\psi(W,\hat{\eta}_{-k};\theta)-\psi(W,\eta;\theta)\\
&=\underbrace{\frac{\hat{q}_{1}(X)_{-k}}{q_1}\psi^a(W,\hat{\eta}_{-k})+\psi^b(W,\hat{\eta}_{-k})\hat{m}_{\tilde{Y}}(X)_{-k}-\frac{q_1(X)}{q_1}\psi^a(W,\eta)-\psi^b(W,\eta)m_{\tilde{Y}}(X)}_{ia}+\underbrace{\left(\psi^b(W,\eta)-\psi^b(W,\hat{\eta}_{-k})\right)}_{ib}\theta.
\end{align*}
Using Assumption \ref{ass:estinfjoint} (i) and (ii) and since
\begin{align*}
&\mathbb{E}\left[\psi(W,\hat{\eta}_{-k};\theta)-\psi(W,\eta;\theta)\right]=\mathbb{E}[ia]+\mathbb{E}[ib]\theta\quad\text{and}\\
&\left\lVert \psi(W,\hat{\eta}_{-k};\theta)-\psi(W,\eta;\theta)\right\rVert_2\leq\lVert ia\rVert_2+\lVert ib\rVert_2\theta,
\end{align*}
Lemma \ref{lm:firststage} implies that $\lVert i \rVert_2=o_p\left(N^{-\frac{1}{2}}\right)$ when $\mathbb{E}[ia]=o_p\left(N^{-\frac{1}{2}}\right)$ and $\lVert ia\rVert_2=o_p(1)$.\\
For a specific $G_{\tau}$ one can write
\begin{align*}
ia_{G_{\tau}}&=\underbrace{\frac{G_{\tau}}{q_1}\left(\frac{\hat{q}_1(X)_{-k}}{\hat{q}_{G_{\tau}}(X)_{-k}}-\frac{q_1(X)}{q_{G_{\tau}}(X)}\right)\left(m_{\tilde{Y}}(G_{\tau}=1,X)-\hat{m}_{\tilde{Y}}(G_{\tau}=1,X)_{-k}\right)}_{ia_{G_{\tau}}1}\\
&+\underbrace{\left(\frac{G_{\tau}}{q_1}\frac{q_1(X)}{q_{G_{\tau}}(X)}-\psi^b(W,\eta)\right)\left(m_{\tilde{Y}}(G_{\tau}=1,X)-\hat{m}_{\tilde{Y}}(G_{\tau}=1,X)_{-k}\right)}_{ia_{G_{\tau}}2}\\
&+\underbrace{\frac{G_{\tau}}{q_1}\left(\frac{\hat{q}_1(X)_{-k}}{\hat{q}_{G_{\tau}}(X)_{-k}}-\frac{q_1(X)}{q_{G_{\tau}}(X)}\right)\left(\tilde{Y}-m_{\tilde{Y}}(G_{\tau}=1,X)\right)}_{ia_{G_{\tau}}3}\\
&+\underbrace{\left(\psi^b(W,\hat{\eta}_{-k})-\psi(W,\eta)\right)\hat{m}_{\tilde{Y}}(G_{\tau}=1,X)_{-k}}_{ia_{G_{\tau}}4}.
\end{align*}
Notice that
\begin{align*}
\mathbb{E}\left[ia_{G_{\tau}}1\right]&\leq\left\lVert\frac{\hat{q}_1(X)_{-k}}{\hat{q}_{G_{\tau}}(X)_{-k}}-\frac{q_1(X)}{q_{G_{\tau}}(X)}\right\rVert_2\times\left\lVert\hat{m}_{\tilde{Y}}(G_{\tau}=1,X)_{-k}-m_{\tilde{Y}}(G_{\tau}=1,X)\right\rVert_2\\
&=o_p\left(N^{-\frac{1}{2}}\right)
\end{align*}
by Assumption \ref{ass:estinfjoint} (v), $\mathbb{E}\left[ia_{G_{\tau}}2\right]=0$ by Assumption \ref{ass:estinfb} (i), $\mathbb{E}\left[ia_{G_{\tau}}3\right]=0$ by the Law of Iterated Expectations and $\mathbb{E}\left[ia_{G_{\tau}}4\right]=o_p\left(N^{-\frac{1}{2}}\right)$ by Assumption \ref{ass:estinfjoint} (iii).\\
Further, $\lVert ia_{G_{\tau}}\rVert_2\leq\lVert ia_{G_{\tau}}1\rVert_2+\lVert ia_{G_{\tau}}2\rVert_2+\lVert ia_{G_{\tau}}3\rVert_2+\lVert ia_{G_{\tau}}4\rVert_2$ and
\begin{align*}
\lVert ia_{G_{\tau}}1\rVert_2&\leq\left\lVert\frac{G_{\tau}}{q_1}\left(\frac{\hat{q}_1(X)_{-k}}{\hat{q}_{G_{\tau}}(X)_{-k}}-\frac{q_1(X)}{q_{G_{\tau}}(X)}\right)\right\rVert_{\infty}\times\left\lVert\hat{m}_{\tilde{Y}}(G_{\tau}=1,X)_{-k}-m_{\tilde{Y}}(G_{\tau}=1,X)\right\rVert_2\\
&\leq C \left\lVert\frac{\hat{q}_1(X)_{-k}}{\hat{q}_{G_{\tau}}(X)_{-k}q_{G_{\tau}}(X)}\right\rVert_{\infty}\times\left\lVert\hat{q}_{G_{\tau}}(X)_{-k}-q_{G_{\tau}}(X)\right\rVert_{\infty}\times\left\lVert\hat{m}_{\tilde{Y}}(G_{\tau}=1,X)_{-k}-m_{\tilde{Y}}(G_{\tau}=1,X)\right\rVert_2\\
&+C\left\lVert\frac{1}{q_{G_{\tau}}(X)}\right\rVert_{\infty}\times\left\lVert\hat{q}_1(X)_{-k}-q_1(X)\right\rVert_{\infty}\times\left\lVert\hat{m}_{\tilde{Y}}(G_{\tau}=1,X)_{-k}-m_{\tilde{Y}}(G_{\tau}=1,X)\right\rVert_2\\
&=o_p(1)
\end{align*}
by Assumptions \ref{ass:estinfp} and \ref{ass:estinfjoint} (iv),
\begin{align*}
\lVert ia_{G_{\tau}}2\rVert_2&\leq\left\lVert\frac{G_{\tau}}{q_1}\frac{q_1(X)}{q_{G_{\tau}}(X)}-\psi^b(W,\eta)\right\rVert_{\infty}\times\left\lVert\hat{m}_{\tilde{Y}}(G_{\tau}=1,X)_{-k}-m_{\tilde{Y}}(G_{\tau}=1,X)\right\rVert_2\\
&=o_p(1)
\end{align*}
by Assumptions \ref{ass:estinfb} (iii) and \ref{ass:estinfjoint} (iv),
\begin{align*}
\lVert ia_{G_{\tau}}3\rVert_2&\leq C\left\lVert\left(\frac{\hat{q}_1(X)_{-k}}{\hat{q}_{G_{\tau}}(X)_{-k}}-\frac{q_1(X)}{q_{G_{\tau}}(X)}\right)G_{\tau}(\tilde{Y}-m_{\tilde{Y}}(G_{\tau}=1,X))\right\rVert_2\\
&=C\left(\mathbb{E}\left[\left(\frac{\hat{q}_1(X)_{-k}}{\hat{q}_{G_{\tau}}(X)_{-k}}-\frac{q_1(X)}{q_{G_{\tau}}(X)}\right)^2G_{\tau}(\tilde{Y}-m_{\tilde{Y}}(G_{\tau}=1,X))^2\right]\right)^{\frac{1}{2}}\\
&\leq C\left\lVert q_{G_{\tau}}(X)\right\rVert_{\infty}\left(\mathbb{E}\left[\left(\frac{\hat{q}_1(X)_{-k}}{\hat{q}_{G_{\tau}}(X)_{-k}}-\frac{q_1(X)}{q_{G_{\tau}}(X)}\right)^2\text{Var}(\tilde{Y}|G_{\tau}=1,X)\right]\right)^{\frac{1}{2}}\\
&\leq C\left\lVert\frac{\hat{q}_1(X)_{-k}}{\hat{q}_{G_{\tau}}(X)_{-k}}-\frac{q_1(X)}{q_{G_{\tau}}(X)}\right\rVert_2\\
&=o_p(1)
\end{align*}
by Assumptions \ref{ass:estinfp}, \ref{ass:estinfy} (ii) and \ref{ass:estinfjoint} (iv), and lastly 
\begin{align*}
\lVert ia_{G_{\tau}}4\rVert_2&\leq\left\lVert\psi^b(W,\hat{\eta}_{-k})-\psi(W,\eta)\right\rVert_{\infty}\times\left\lVert\hat{m}_{\tilde{Y}}(G_{\tau}=1,X)_{-k}-m_{\tilde{Y}}(G_{\tau}=1,X)\right\rVert_2\\
&+\left\lVert\left(\psi^b(W,\hat{\eta}_{-k})-\psi^b(W,\eta)\right)m_{\tilde{Y}}(G_{\tau}=1,X)\right\rVert_2\\
&=o_p(1)
\end{align*}
by Assumptions \ref{ass:estinfjoint} (i), (iii) and (iv).\\~\\
\textit{Bounding ii}
\begin{align*}
\lvert ii\rvert&\leq\left\lvert1-\frac{1}{N}\sum_{k=1}^K\sum_{i\in\mathcal{I}^k}^n\psi^b(W_i,\eta)\right\rvert\times\left\lvert\left(1+\left(\frac{1}{N}\sum_{k=1}^K\sum_{i\in\mathcal{I}^k}^n\psi^b(W_i,\eta)-1\right)\right)^{-1}\right\rvert\\
&=O_p\left(N^{-\frac{1}{2}}\right)\left(1+o_p(1)\right)^{-1}\\
&=O_p\left(N^{-\frac{1}{2}}\right)
\end{align*}
since by a weak law of large numbers sample means should converge against their expectations and $\mathbb{E}\left[\psi^b(W,\eta)\right]=1$ by Assumption \ref{ass:estinfb} (i).\\~\\
\textit{Bounding iii}
\begin{align*}
\lvert iii\rvert&\leq\left\lvert\underbrace{\frac{1}{\frac{1}{N}\sum_{k=1}^K\sum_{i\in\mathcal{I}^k}^n\psi^b(W_i,\hat{\eta}_{-k})}}_{iiia}\right\rvert\times\left\lvert\underbrace{\frac{1}{\frac{1}{N}\sum_{k=1}^K\sum_{i\in\mathcal{I}^k}^n\psi^b(W_i,\eta)}}_{iiib}\right\rvert\\
&\times\left\lvert\underbrace{\frac{1}{N}\sum_{k=1}^K\sum_{i\in\mathcal{I}^k}^n\psi^b(W_i,\eta)-\frac{1}{N}\sum_{k=1}^K\sum_{i\in\mathcal{I}^k}^n\psi^b(W_i,\hat{\eta}_{-k})}_{iiic}\right\rvert
\end{align*}
where $\lvert iiia\rvert\leq\lVert iiia\rVert_{\infty}\leq C$, $\lvert iiib\rvert\leq\lVert iiib\rVert_{\infty}\leq C$ by Assumption \ref{ass:estinfb} (ii). Further, conditions in Assumption \ref{ass:estinfjoint} (i) and (ii) in combination with Lemmma \ref{lm:firststage} implies that $\lvert iiic\rvert=o_p\left(N^{-\frac{1}{2}}\right)$ with high probability. Therefore $\lvert iii\rvert=O_p(1)O_p(1)o_p\left(N^{-\frac{1}{2}}\right)=o_p\left(N^{-\frac{1}{2}}\right)$.
\subsection{Proof of Corollary \ref{cor:estcs}}\label{app:estcs}
$\mathbb{E}\left[\psi^*_{CS}(W,\eta;\theta)\right]=0$ using Assumptions \ref{ass:idcs} and the respective condition in settings (CS-1)-(CS-5). For all $\psi^{*b}_{CS}(W,\eta)$ it is easy to see that the conditions in Assumption \ref{ass:estinfb} are satisfied. Therefore it remains to show that the conditions in Assumptions \ref{ass:estinfjoint} hold under the convergence conditions stated in the corollary.
\begin{itemize}
\item[(a)] For $\psi^{*b}_{CS-1}(W,\eta)=\frac{DT}{p_{DT}}$ the conditions in Assumption \ref{ass:estinfjoint} (i)-(iii) are trivially satisfied. Notice that $m_Y(1,1,X)$ is redundant and that for all $(d,t)\in\{(0,1),(1,0),(0,0)\}$
\begin{align*}
&\left\lVert\frac{\hat{p}_{D=1,T=1}(X)_{-k}}{\hat{p}_{D=d,T=t}(X)_{-k}}-\frac{p_{D=1,T=1}(X)}{p_{D=d,T=t}(X)}\right\rVert_2\\
&\leq\left\lVert\frac{1}{\hat{p}_{D=d,T=t}(X)_{-k}}\right\rVert_{\infty}\times\left\lVert\hat{p}_{D=1,T=1}(X)_{-k}-p_{D=1,T=1}(X)\right\rVert_2\\
&+\left\lVert\frac{p_{D=1,T=1}(X)}{\hat{p}_{D=d,T=t}(X)_{-k}p_{D=d,T=t}(X)}\right\rVert_{\infty}\times\left\lVert\hat{p}_{D=d,T=t}(X)_{-k}-p_{D=d,T=t}(X)\right\rVert_2\\
&\leq C\times\left(\epsilon_{p_{D=1,T=1}(X)}+\epsilon_{p_{D=d,T=t}(X)}\right)
\end{align*}
by Assumption \ref{ass:estinfp}. This immediately implies that Assumptions \ref{ass:estinfjoint} (iv) and (v) are satisfied under the conditions stated. 
\item[(b)] For $\psi^{*b}_{CS-2}(W,\eta)=\frac{Dp_T(X)+Tp_D(X)-p_D(X)p_T(X)}{p_{DT}}$ write
\begin{align*}
&\psi^{*b}_{CS-2}(W,\hat{\eta}_{-k})-\psi^{*b}_{CS-2}(W,\eta)\\
&=\frac{1}{p_{DT}}\left(D-p_D(X)\right)\left(\hat{p}_T(X)_{-k}-p_T(X)\right)+\frac{1}{p_{DT}}\left(T-p_T(X)\right)\left(\hat{p}_D(X)_{-k}-p_D(X)\right)\\
&+\frac{1}{p_{DT}}\left(\hat{p}_T(X)_{-k}-p_T(X)\right)\left(p_D(X)-\hat{p}_D(X)_{-k}\right).
\end{align*}
Then 
$\left\lVert\psi^{*b}_{CS-2}(W,\hat{\eta}_{-k})-\psi^{*b}_{CS-2}(W,\eta)\right\rVert_2\leq\epsilon_{p_T(X)}+\epsilon_{p_D(X)}=o_p(1)$ by Assumption \ref{ass:estinfp} and the convergence conditions stated. Similarly, $\left\lVert\psi^{*b}_{CS-2}(W,\hat{\eta}_{-k})-\psi^{*b}_{CS-2}(W,\eta)\right\rVert_{\infty}\leq\left\lVert\hat{p}_T(X)_{-k}-p_T(X)\right\rVert_{\infty}\times\left\lVert\hat{p}_D(X)_{-k}-p_D(X)\right\rVert_{\infty}=O_p(1)$ by Assumption \ref{ass:estinfp}. This verifies Assumption \ref{ass:estinfjoint} (i).\\
Also $\mathbb{E}\left[\psi^{*b}_{CS-2}(W,\hat{\eta}_{-k})-\psi^{*b}_{CS-2}(W,\eta)\right]\leq\epsilon_{p_T(X)}\times\epsilon_{p_D(X)}=o_p\left(N^{-\frac{1}{2}\frac{r}{r-1}}\right)$
which verifies Assumption \ref{ass:estinfjoint} (ii).\\
To verify Assumption \ref{ass:estinfjoint} (iii), notice that for $r>2$
\begin{align*}
&\left\lVert\left(\hat{p}_T(X)_{-k}-p_T(X)\right)\left(p_D(X)-\hat{p}_D(X)_{-k}\right)\right\rVert_{\frac{r}{r-1}}\\
&=\left(\mathbb{E}\left[\left(\hat{p}_T(X)_{-k}-p_T(X)\right)^{\frac{r}{r-1}}\left(p_D(X)-\hat{p}_D(X)_{-k}\right)^{\frac{r}{r-1}}\right]\right)^\frac{r-1}{r}\\
&\leq\left\lVert\hat{p}_D(X)_{-k}-p_D(X)\right\rVert_{\infty}^{\frac{1}{r}}\times\left\lVert\hat{p}_T(X)_{-k}-p_T(X)\right\rVert_{\infty}^{\frac{1}{r}}\times\left(\epsilon_{p_T(X)}\times\epsilon_{p_D(X)}\right)^\frac{r-1}{r}\\
&=o_p\left(N^{-\frac{1}{2}}\right).
\end{align*}
Then for any $d,t\in\{0,1\}$
\begin{align*}
&\mathbb{E}\left[\left(\psi^{*b}_{CS-2}(W,\hat{\eta}_{-k})-\psi^{*b}_{CS-2}(W,\eta)\right)\left(\hat{m}_Y(d,t,X)_{-k}-m_Y(d,t,X)\right)\right]\\
&\leq\left\lVert\left(\hat{p}_T(X)_{-k}-p_T(X)\right)\left(p_D(X)-\hat{p}_D(X)_{-k}\right)\right\rVert_{\frac{r}{r-1}}\times\left\lVert\hat{m}_Y(d,t,X)_{-k}-m_Y(d,t,X)\right\rVert_r\\
&=o_p\left(N^{-\frac{1}{2}}\right),\\
&\mathbb{E}\left[\left(\psi^{*b}_{CS-2}(W,\hat{\eta}_{-k})-\psi^{*b}_{CS-2}(W,\eta)\right)m_Y(d,t,X)\right]\\
&\leq\left\lVert\left(\hat{p}_T(X)_{-k}-p_T(X)\right)\left(p_D(X)-\hat{p}_D(X)_{-k}\right)\right\rVert_{\frac{r}{r-1}}\times\left\lVert m_Y(d,t,X)\right\rVert_r\\
&=o_p\left(N^{-\frac{1}{2}}\right)
\end{align*}
under the conditions stated in the corollary. Further,
\begin{align*}
\left\lVert\left(D-p_D(X)\right)\left(\hat{p}_T(X)_{-k}-p_T(X)\right)\right\rVert_{\frac{2r}{r-2}}&\leq\left\lVert\hat{p}_T(X)_{-k}-p_T(X)\right\rVert_{\frac{2r}{r-2}}\\
&\leq\left\lVert\hat{p}_T(X)_{-k}-p_T(X)\right\rVert_{\infty}^{\frac{r+2}{2r}}\times\left\lVert\hat{p}_T(X)_{-k}-p_T(X)\right\rVert_2^{\frac{r-2}{2r}}\\
&=o_p(1),
\end{align*}
similarly $\left\lVert\left(T-p_T(X)\right)\left(\hat{p}_D(X)_{-k}-p_D(X)\right)\right\rVert_{\frac{2r}{r-2}}=o_p(1)$ and
\begin{align*}
&\left\lVert\left(\hat{p}_T(X)_{-k}-p_T(X)\right)\left(p_D(X)-\hat{p}_D(X)_{-k}\right)\right\rVert_{\frac{2r}{r-2}}\\
&\leq\left\lVert\hat{p}_T(X)_{-k}-p_T(X)\right\rVert_{\infty}^{\frac{r+2}{2r}}\times\left\lVert\hat{p}_D(X)_{-k}-p_D(X)\right\rVert_{\infty}^{\frac{r+2}{2r}}\times\left(\epsilon_{p_T(X)}\times\epsilon_{p_D(X)}\right)^{\frac{r-2}{2r}}\\
&=o_p(1)
\end{align*}
by Assumption \ref{ass:estinfp}, the fact that $r>2$ and the convergence conditions stated. Then for all $d,t\in\{0,1\}$
\begin{align*}
\left\lVert\left(\psi^{*b}_{CS-2}(W,\hat{\eta}_{-k})-\psi^{*b}_{CS-2}(W,\eta)\right)m_Y(d,t,X)\right\rVert_2&\leq\left\lVert\psi^{*b}_{CS-2}(W,\hat{\eta}_{-k})-\psi^{*b}_{CS-2}(W,\eta)\right\rVert_{\frac{2r}{r-2}}\times\left\lVert m_Y(d,t,X)\right\rVert_{r}\\
&=o_p(1).
\end{align*}
Assumption \ref{ass:estinfjoint} (iv) and (v) directly follow from the conditions stated.
\item[(c)] For $\psi^{*b}_{CS-3}(W,\eta)=\frac{T\left(D-p_D(1,X)\right)}{p_{D}(1)p_T}+\frac{p_D(1,X)}{p_D(1)}$ write
\begin{align*}
\psi^{*b}_{CS-3}(W,\hat{\eta}_{-k})-\psi^{*b}_{CS-3}(W,\eta)=\frac{1}{p_D(1)p_T}\left(\left(p_T-T\right)\left(\hat{p}_{D}(1,X)_{-k}-p_D(1,X)\right)\right).
\end{align*}
Since 
\begin{align*}
&\left\lVert\psi^{*b}_{CS-3}(W,\hat{\eta}_{-k})-\psi^{*b}_{CS-3}(W,\eta)\right\rVert_2\leq\epsilon_{p_D(1,X)}=o_p(1)\quad\text{and}\\
&\left\lVert\psi^{*b}_{CS-3}(W,\hat{\eta}_{-k})-\psi^{*b}_{CS-3}(W,\eta)\right\rVert_{\infty}\leq\left\lVert\hat{p}_{D}(1,X)_{-k}-p_D(1,X)\right\rVert_{\infty}=O_p(1)
\end{align*}
Assumption \ref{ass:estinfjoint} (i) is verified.\\
Assumptions \ref{ass:estinfjoint} (ii) holds since trivially $\mathbb{E}\left[\psi^{*b}_{CS-3}(W,\hat{\eta}_{-k})-\psi^{*b}_{CS-3}(W,\eta)\right]=0$ by the Law of Iterated Expectations. Similarly, for all $d,t\in\{0,1\}$ the first part of Assumption \ref{ass:estinfjoint} (iii) is verified by
$\mathbb{E}\left[\left(\psi^{*b}_{CS-3}(W,\eta)-\psi^{*b}_{CS-3}(W,\hat{\eta}_{-k})\right)\hat{m}_Y(d,t,X)_{-k}\right]=0$.
For the second part of Assumption \ref{ass:estinfjoint} (iii) write
\begin{align*}
\left\lVert\left(\psi^{*b}_{CS-3}(W,\eta)-\psi^{*b}_{CS-3}(W,\hat{\eta}_{-k})\right)m_Y(d,t,X)\right\rVert_2&\leq\left\lVert\hat{p}_{D}(1,X)_{-k}-p_D(1,X)\right\rVert_{\frac{2r}{r-2}}\times\left\lVert m_Y(d,t,X)\right\rVert_{r}\\
&\leq\left\lVert\hat{p}_{D}(1,X)_{-k}-p_D(1,X)\right\rVert_{\infty}^{\frac{r+2}{2r}}\times\left(\epsilon_{p_D(1,X)}\right)^\frac{r-2}{2r}\\
&=o_p(1)
\end{align*}
for all $d,t\in\{0,1\}$ using Assumption \ref{ass:estinfp}, the fact that $r>2$ and the convergence conditions stated.\\
Assumptions \ref{ass:estinfjoint} (iv) and (v) directly follow from the conditions stated.
\item[(d)] For $\psi^{*b}_{CS-4}(W,\eta)=\frac{D}{p_{D}}$ the conditions in Assumption \ref{ass:estinfjoint} (i)-(iii) are trivially satisfied. Assumptions \ref{ass:estinfjoint} (iv) and (v) directly follow from the conditions stated.
\item[(e)] For $\psi^{*b}_{CS-5}(W,\eta)=1$ the conditions in Assumption \ref{ass:estinfjoint} (i)-(iii) are trivially satisfied. Assumptions \ref{ass:estinfjoint} (iv) and (v) directly follow from the conditions stated.
\end{itemize}
\subsection{Proof of Corollary \ref{cor:estpa}}\label{app:estpa}
$\mathbb{E}\left[\psi^*_{PA}(W,\eta;\theta)\right]=0$ using Assumptions \ref{ass:idpa} and the respective conditions (PA-1) or (PA-2). For all $\psi^{*b}_{PA}(W,\eta)$ it is easy to see that the conditions in Assumption \ref{ass:estinfb} and Assumption \ref{ass:estinfjoint} (i)-(iii) are satisfied. Assumptions \ref{ass:estinfjoint} (iv) and (v) directly follow from the conditions stated in the corollary.
\subsection{Proof of Corollary \ref{cor:redcs2}}\label{app:redcs2}
Assumption \ref{ass:estinfb} and Assumption \ref{ass:estinfjoint} (i)-(iii) are trivially satisfied when using $\psi'_{CS-2}(W;\theta)$. Notice that $m_Y(1,1,X)$ is redundant. Assumptions \ref{ass:estinfjoint} (iv) and (v) then directly follow from the conditions stated in the corollary.
\subsection{Proof of Corollary \ref{cor:redcs4}}\label{app:redcs4}
Assumption \ref{ass:estinfb} and Assumption \ref{ass:estinfjoint} (i)-(iii) follow similarly to the proof of Corollary \ref{cor:estcs} for $(d,t)\in\{(0,1),(0,0)\}$. Notice that $m_Y(1,1,X)$ and $m_Y(1,0,X)$ are redundant. Assumption \ref{ass:estinfb} and Assumption \ref{ass:estinfjoint} (i)-(iii) are then trivially satisfied for $(d,t)\in\{(1,1),(1,0)\}$. Since $m_Y(1,1,X)$ and $m_Y(1,0,X)$ are redundant, Assumptions \ref{ass:estinfjoint} (iv) and (v) then directly follow from the conditions stated in the corollary.
\end{appendix}
\end{document}